\documentclass[sn-mathphys-num]{sn-jnl}

\usepackage{graphicx}%
\usepackage{multirow}%
\usepackage{amsmath,amssymb,amsfonts}%
\usepackage{amsthm}%
\usepackage{mathrsfs}%
\usepackage[title]{appendix}%
\usepackage{xcolor}%
\usepackage{textcomp}%
\usepackage{manyfoot}%
\usepackage{booktabs}%
\usepackage{algorithm}%
\usepackage{algorithmicx}%
\usepackage{algpseudocode}%
\usepackage{listings}%

\usepackage{mathcomp}
\usepackage{comment}
\usepackage{subcaption}
\usepackage{url}

\usepackage{eurosym}             
\usepackage[export]{adjustbox}   

\usepackage{lmodern}
\usepackage{fix-cm}
\usepackage[T1]{fontenc}
\usepackage{textcomp}
\usepackage{newtxtext,newtxmath}
\usepackage{todonotes}

\usepackage{hyperref}

\begin{document}

\title{Spatio-temporal characterisation of underwater noise through semantic trajectories}


\author*[1]{\fnm{Giulia} \sur{Rovinelli}}\email{giulia.rovinelli@unive.it}

\author[2]{\fnm{Davide} \sur{Rocchesso}}\email{davide.rocchesso@unimi.it}

\author[1,3]{\fnm{Marta} \sur{Simeoni}}\email{simeoni@unive.it}
\author[4]{\fnm{Esteban} \sur{Zim\'anyi}}\email{esteban.zimanyi@ulb.be}
\author[1]{\fnm{Alessandra} \sur{Raffaet\`a}}\email{raffaeta@unive.it}

\affil*[1]{\orgname{Ca' Foscari University of Venice}, \orgaddress{\city{Venice}, \country{Italy}}}

\affil[2]{\orgname{Universit\`a degli studi di Milano Statale}, \orgaddress{\city{Milano}, \country{Italy}}}

\affil[3]{\orgname{European Centre for Living Technology (ECLT)}, \orgaddress{\city{Venice}, \country{Italy}}}

\affil[4]{\orgname{Universit\'e Libre de Bruxelles}, \orgaddress{\city{Brussels}, \country{Belgium}}}


\abstract{Underwater noise pollution from human activities, particularly shipping, has been recognised as a serious threat to marine life. The sound generated by vessels can have various adverse effects on fish and aquatic ecosystems in general. In this setting, the estimation and analysis of the underwater noise produced by vessels is an important challenge for the preservation of the marine environment. 
In this paper we propose a model for the spatio-temporal characterisation of the underwater noise generated by vessels. The approach is based on the reconstruction of the vessels' trajectories from 
Automatic Identification System (AIS) data and on their deployment in a spatio-temporal database. Trajectories are enriched with semantic information like the acoustic characteristics of the vessels' engines or the activity performed by the vessels. We define a model for underwater noise propagation and use the trajectories' information to infer how noise propagates in the area of interest. 
We develop our approach for the case study of the fishery activities in the Northern Adriatic sea, an area of the Mediterranean sea which is well known to be highly exploited. 
We implement our approach using MobilityDB, an open source geospatial trajectory data management and analysis platform, which offers spatio-temporal operators and indexes improving the efficiency of our system. 
We use this platform to conduct various analyses of the underwater noise generated in the Northern Adriatic Sea, aiming at estimating the impact of fishing activities on underwater noise pollution and at demonstrating the flexibility and expressiveness of our approach. 
}


\maketitle

\section{Introduction}\label{introduction}
Underwater noise generated by human activities, especially from shipping, is known to produce short- and long-term effects on marine animal species. This noise pollution can disrupt the natural acoustic environment, leading to several adverse consequences.
Some of the negative impacts include interference with communication, changes in behavior, stranding, and increased mortality rates~\cite{slabbekoorn2010noisy,williams2015impacts}.
Therefore, characterising underwater noise in a specific area is crucial for monitoring the health of aquatic life, assessing potential risks, and providing valuable information to ecologists and policymakers. This enables the development of effective strategies to maintain a productive and healthy ecosystem.

However, measuring underwater noise is a complex and resource-intensive task. It necessitates the use of hydrophones (underwater microphones) and requires a team of experts for deployment and calibration. Once the data has been collected, it must be processed and analysed to extract valuable insights. Additionally, there is the need to estimate underwater noise in areas where data collection is impractical or to extend coverage beyond what the hydrophones can monitor.
Many approaches in the literature rely on acoustic models generated through numerical simulations to predict sound levels in the target area (see e.g. \cite{MacGillivray2014,Larayedh2024,ghezzo2024natural}). These models simulate sound propagation by accounting for reflection, diffraction, and absorption phenomena.  They require precise input data and significant computational resources to create accurate acoustic maps. Typically, these models consider various environmental variables (like sea temperature, wind, waves, and salinity), detailed bathymetric information, and sound speed profiles at different depths. They may also incorporate Automatic Identification System (AIS) data to track vessel movements and include their noise emissions in the simulations.

In this paper, we follow a complementary approach that avoids using numerical simulation and fully relies on AIS data to calculate the underwater noise generated by  vessels in space and time. Similar approaches in the literature are e.g. Erbe et al.~\cite{erbe2012mapping} and Neenan et al.~\cite{neenan2016modeling}, where the authors' main goal is to readily provide noise maps of the area of interest.
However, rather than developing an ad-hoc analysis for a specific case study, our aim is to propose a conceptual framework for underwater noise characterisation that can be easily instantiated on any sea area and can be used as a quick and effective means to monitor the noise pollution.

Our  framework is based on semantic trajectories~\cite{parent2013Semantic, mello2019master}. Starting from AIS data, we reconstruct the vessels trajectories and deploy them in a spatio-temporal database.  These trajectories are enhanced with semantic information, such as the acoustic characteristics of the vessels' engines and the activities conducted along their paths, which are then used to infer how the noise spreads in the area of interest.

To showcase the potential of the approach we consider the fishing activities of the Northern Adriatic Sea, which is known to be one of the most exploited areas of the Mediterranean Sea, so the underwater noise pollution is certainly among the effects of the intensive fishery activity.
We build on our previous work \cite{rovinelli2021multiple,brandoli2022multiple}, which describes and implements a spatio-temporal database of the fishing activities in the Northern Adriatic Sea, and on a preliminary underwater noise model presented in Rovinelli et al.~\cite{rovinelli2024using}. The trajectories of the fishing vessels are reconstructed starting from the AIS data, sent by ships and received by ground stations on the Italian coast. The dataset considered in this paper comprises all AIS data of the Italian and Croatian fishing vessels for the year 2020. In order to determine the acoustic characteristics of the vessels' engines and fine tune the propagation model, we  take advantage of the direct acoustic measurements produced by the Interreg project SOUNDSCAPE~\cite{SoundscapeProj} that carried out an acoustic monitoring in the Northern Adriatic Sea from March 2020 to June 2021.

To estimate the generated noise, we define a model for underwater sound propagation and instantiate it by considering four different frequencies.  We present the theoretical laws 
governing the underwater sound propagation and we also describe how we obtain the estimation for source sound levels, propagation loss and ambient noise, necessary for the definition of the model. We then use the spatio-temporal database to calculate the noise produced along the vessels trajectories. The sea area is partitioned into a regular grid with each cell measuring 1 km by 1 km, 
allowing for collecting noise intensity data in both space and time. Instead of using physical listening points (hydrophones), we treat the centroid of each cell as a \emph{virtual} listening point where we calculate the received sound pressure level.

The conceptual framework has been implemented in MobilityDB \cite{MobilityDBTODS2020}, an open-source platform for managing and analysing geospatial trajectory data. We use this platform to conduct various analyses that aim at estimating the impact of fishing activities on underwater noise pollution 
and at demonstrating the flexibility and expressivity of our approach. We first investigate the underwater noise generated by fishing activities at different frequencies. Then we look into the effects of the COVID-19 in the Northern Adriatic Sea, by comparing the underwater sound pressure levels in two periods, April and June 2020, i.e., during, and after the lockdown imposed in Italy. Finally, we show that our tool  allows for the visualisation of the underwater noise dynamics for a set of vessels chosen by the user according to several criteria. 

In summary, the main contributions of this work are the following: 
\begin{itemize}
\item We define a model for underwater noise propagation that refines the preliminary version presented in Rovinelli et al.~\cite{rovinelli2024using}.  First, in the computation of the source level of ships we have added the contribution of speed. In fact, as noted in~\cite{chion2019meta}, speed can influence the broadband source level of ships. Second, the transmission loss is no longer modelled simply by a spherical spreading law, but it is implemented as a combination between spherical propagation and mode stripping following Ainslie's model~\cite{ainslie2010principles}. This allows us to take into consideration the bathymetry of the study basin.
\item The model is instantiated by considering four different frequencies: 63~Hz and 125~Hz,  established as standard frequencies by the \emph{European Marine Strategy Framework Directive} (MSFD), and 400~Hz and 4000~Hz. The latter two frequencies allow us to show the increasing significance of environmental factors at higher ranges. For each frequency we need to calibrate the model parameters, specifically the source level of the ships, the ambient noise, the increment due to fishing and the transition range between spherical and mode stripping. 
\item We define a conceptual framework for the spatio-temporal characterisation of underwater noise based on semantic trajectories. The movement of the vessels is reconstructed from AIS data and enriched with semantic information which allows for the estimation of the noise generated by the vessels. To calculate the noise level induced by the vessels we propose an algorithm which has been improved w.r.t. the one in Rovinelli et al.~\cite{rovinelli2024using} to consider the refined model of noise propagation and the different frequencies.
\item We implement the framework in MobilityDB  to create a spatio-temporal database of the Northern Adriatic Sea. 
The implementation has been enhanced by extensively exploiting the temporal types and spatio-temporal operations offered by MobilityDB. The result is a more compact and efficient code, leading to much better execution times. 
\item We conduct several analyses illustrating the impact of the fishing activities on underwater noise pollution in the Northern Adriatic Sea. These are only a few examples of the analyses that can be performed using the spatio-temporal database we have created. 
\end{itemize}

The paper is organised as follows. Section~\ref{sec:RelatedWork} presents the literature on underwater noise characterisation. 
Section~\ref{sec:UnderwaterModel} describes our model for the underwater sound propagation.  Section~\ref{sec:modelmplementation} outlines the trajectory reconstruction and enrichment, the creation of a spatio-temporal database by means of MobilityDB and the description of the algorithm computing the underwater noise maps. 
Section~\ref{sec: results} presents some analyses conducted on the obtained spatio-temporal database and some notes on model validation. 
Finally, some closing remarks are outlined in Section~\ref{sec: conclusion}.

\section{Related work}\label{sec:RelatedWork}
Underwater noise arising from human activities is known to have various adverse effects on aquatic life. These can range from acute effects such as permanent or temporary hearing impairment to chronic effects such as developmental deficiencies and physiological stress~\cite{slabbekoorn2010noisy,williams2015impacts}.
As a general study, we mention the work by Cruz et al.~\cite{cruz2021study} that summarised the status of European waters regarding continuous underwater radiated noise from shipping. The goal was to provide recommendations on possible future activities. 
The work focused on four main topics: characteristics and quantification of noise sources from various ship types, impacts on marine fauna, existing policies, including guidelines, decisions, resolutions and regulations and mitigation measures for the abatement of ship noise and noise-related impact. 

The specific topic of interest in this paper, i.e. underwater noise generated by (fishing) vessels, has been explored by various works in the literature and we illustrate below the main contributions. We proceed by first discussing some approaches dealing with direct underwater noise measurements and then by considering models of underwater noise characterisation,  based both on numerical simulation and on AIS data only.

\subsection{Underwater noise characterisation by direct measurements}
\label{sec:RelatedWorkMeasurements}
Some works in the literature address the issue of underwater noise data acquisition, transmission and storage. 
The most interesting one in our perspective is the dataset produced by the SOUNDSCAPE  project~\cite{SoundscapeProj}, which carried out an acoustic monitoring of the Northern Adriatic Sea from March 2020 to June 2021. Nine monitoring stations were set up encompassing different environmental characteristics. Two datasets have been realised composed of 20 and 60 seconds averaged Sound Pressure Levels (SPLs) data in a wide range of frequencies recorded at the nine stations. These datasets are available on Zenodo ~\cite{ZenodoPicciulin} and the whole process of data acquisition, storing and post-processing is described by Petrizzo et al. in~\cite{picciulin2023data}. 
Furthermore, Picciulin et al. ~\cite{picciulin2023SR} perform some analyses and describe the spatial and temporal variations of the sound pressure levels recorded by the SOUNDSCAPE project over one year (March 1, 2020 - March 31, 2021).

Other approaches propose methods for acoustic real-time measurements and monitoring, such as Diviacco et al.~\cite{diviacco2021underwater} and Moran et al.~\cite{moran2019multi}. Differently, Farcas et al.~\cite{farcas2020validated} carried out multi-site measurements to validate a large-scale shipping noise map constructed using a generic shipping noise model.
Finally, the work by Picciulin et al. ~\cite{picciulin2021not} shows two acoustic surveys conducted at 40 listening points distributed along the three inlets that connect the Venice lagoon to the sea, in order to characterise the local noise levels and evaluate the fish spatial distribution by means of its sounds.

\subsection{Approaches based on numerical simulation}
Most of the approaches to underwater noise characterisation are based on acoustic models obtained by numerical simulation to predict the sound levels in the area of interest. They simulate  sound propagation taking into consideration reflection, diffraction and absorption phenomena and require precise information in input and a relevant computational effort (in terms of time and resources) to produce the acoustic maps of the area of interest. Such models usually account for many environmental variables (e.g. sea temperature, wind, waves, salinity), precise bathymetry information and sound speed profiles, possibly at different sea depths. They use AIS data to derive the vessels' trajectories and include their emitted sound into the model. The results obtained are often validated through real acoustic data measured on specific sites. 

As a first approach in this category, we mention the work by MacGillivray et al.~\cite{MacGillivray2014}. The authors present an acoustic model to predict anthropogenic sound levels at the Great Barrier Reef Marine Park in Australia. The model uses AIS data and wind speed data to simulate the time-dependent noise field in the area of interest and for the frequencies $64~\rm{kHz}$ and $375~\rm{kHz}$.
The model uses three months of AIS data and assign acoustic source levels to each vessel according to predefined values already calculated for the various vessels categories. The model includes environmental parameters such as ocean temperature and salinity as well as bathymetry information. The obtained results were compared against  real data collected by an acoustic recorder placed on a specific site in the same period.

A similar approach is used by Larayedh et al.~\cite{Larayedh2024} to investigate the shipping noise in the Red Sea for the frequency band 40-100~Hz. The acoustic model, based on simulation, includes two months of AIS data belonging to specific categories of shipping vessels (tankers, bulkers, containser ships, open hatch vessels and vehicle carriers), whose sound levels were derived from~\cite{MacGillivray2021}. Bathymetry data, sea temperature and salinity are  included, as well as spatial and temporal sound speed profiles specifically calculated for the Red Sea. The authors emphasise the computational effort needed to calculate the resulting maps of predicted noise levels, which required the use of a supercomputer. Results include the maps of predicted spectral noise level (averaged over the considered frequency band) for a specific day at different depths, and maps considering the two months under exam.
 
Along the same line, Ghezzo et al. present in~\cite{ghezzo2024natural} a numerical reconstruction of the sound field in the Northern Adriatic sea basin for the year 2020. Acoustic modelling was performed by the Quonops\textcopyright~\cite{Quonops} underwater noise prediction system, which is based on simulation. \textit{Natural Sound Maps} were produced by taking into account the sound propagation properties of the local environment (hourly wind and waves data, daily mean sea temperature and salinity, bathymetry data). Additionally, a combination of natural sources and AIS-based marine traffic were used to produce the \textit{Baseline sound maps}. 
The AIS dataset included all types of vessels and  the source level noise produced by the various categories of vessels were derived from~\cite{MacGillivray2021}. The trawling activity of the fishing vessels was taken into account to consider an increased  noise level when the trawler is in use. 
Calibration of the AIS-based model was performed using the SOUNDSCAPE~\cite{SoundscapeProj} measured data.
The authors present annual Baseline sound maps and excess 20~dB sound maps for the frequencies 63~Hz, 125~Hz and 250~Hz. The study included specific analyses on protected  areas of the Northern Adriatic sea.

The work by Ghezzo et al.~\cite{ghezzo2024natural} is very important in our setting because it considers the same area (the Northern Adriatic sea), period (2020), source of ground truth data (the SOUDSCAPE time series measurements) and some common frequencies (63~Hz and 125~Hz). However, the results of the two studies cannot be directly compared because Ghezzo et al. consider the AIS data of all vessels while we focus on fishing vessels only, so the majority of the marine traffic is not included in our framework.

\subsection{Approaches centered on AIS data}
Other proposals in the literature avoid the use of numerical simulation and give a more simplified description of the sound propagation effects. They are based on AIS data and they usually adopt a sound propagation model to calculate the transmission loss in the area of interest.
The goal is to readily provide acoustic maps that can be used by policy makers for a first quantitative spatio-temporal underwater noise evaluation. They are computationally efficient, even for large datasets.

As a first approach in this setting, we mention the one by Erbe et al.~\cite{erbe2012mapping} that uses the ship transits derived from AIS data and calculated by the Canadian Guard Coast. The authors  apply a sound propagation model to derive a cumulative large-scale noise map of the area of interest for the frequency band 10-2000~Hz. The authors include a comparison with field measurements.
Similarly, Neenan et al.~\cite{neenan2016modeling} develop a vessel noise modelling method using AIS data and online data on estimated source levels of individual ships. The authors divide the area of interest into a spatial grid of $1\times1$~km and consider a single frequency of 80~Hz. They calculate the propagation loss in a 5~km ray around each vessel's position taking into account also sediment type and bathymetry. The goal is to produce heat-maps of average received sound levels over monthly periods.   

The approach followed in this paper is similar to that in ~\cite{erbe2012mapping,neenan2016modeling} since we also use 
AIS data and a model of sound propagation to derive a spatio-temporal characterisation of the underwater noise of the area of interest. However, rather than developing an ad-hoc analysis for a specific case study, our aim is to exploit the AIS based approach to develop a general framework for underwater noise characterisation, to be used as a quick and effective means to monitor the area of interest. The framework is based on the construction of semantic trajectories of the fishing vessels and on the implementation of a spatio-temporal database. We use the information on the engine power of the fishing vessels to determine the source noise levels generated by the vessels and then exploit their trajectories to deduce how such noise propagates. Having semantic information allows us to distinguish between different behaviours of the fishing vessels, with the possibility of identifying the moments in which they are fishing, an activity which increases the generated noise.
The spatio-temporal database can be updated daily in a few minutes and can be queried to answer any question about underwater noise with varying temporal and spatial granularity (from one minute per $1\times1$~km cell to an arbitrary period of time and area of interest), and with the possibility of aggregating on an arbitrary set of boats. For instance, a user may request to visualise the noise produced by a single vessel along its trajectory, considering also the increase of noise while fishing, or to show noise maps for a chosen area and time period, or to produce videos illustrating the underwater noise dynamics in timelapse.  

A further proposal in this setting that demonstrates the potential of using AIS data for underwater noise estimation is Jallkanen et al. in~\cite{JALKANEN2022}. This work uses worldwide terrestrial and satellite AIS data covering the years 2014-2020 and the STEAM (Ship trafic Abatement Model) tool by the Finnish Meteorological Institute~\cite{JOHANSSON2017} to estimate vessel noise source levels. The STEAM model predicts instantaneous vessel noise levels that were then cumulated over time to produce a noise source energy map. 
The instantaneous noise levels in each vessel position (and in a sphere around the source) were reported in a spatio-temporal grid cell, whose resolution is 0.1$^{\rm{o}}$ (WGS84 coordinate system). The gridded data were produced for the frequencies 63~Hz, 125~Hz and 2000~Hz. Results included a quantification of the global  underwater noise emissions from shipping in the period 2014-2020 (showing that at the current rate the emissions are doubling every 11.5 years), a study of the emissions in different areas (showing a large variability of shipping noise in different regions), a study on the emissions during the COVID19 pandemic (showing a drastic reduction of emission levels globally) and a study of the emissions of the various vessels categories (indicating that container ships are the largest contributor to shipping noise emissions).

\section{Underwater noise model}
\label{sec:UnderwaterModel}
In this section we describe a model for underwater sound propagation that is used in Section~\ref{sec:modelmplementation} to provide a spatio-temporal characterisation of underwater noise in the Northern Adriatic Sea. 
We present the theoretical laws 
governing the underwater sound propagation and we also describe how we obtain the estimation for source sound levels, propagation loss and ambient noise, necessary for the definition of the model. 

The estimation is based on real measurements. We use the dataset containing the 60 seconds SPL in different frequencies delivered by the project SOUNDSCAPE~\cite{SoundscapeProj,ZenodoPicciulin}, which carried out an acoustic monitoring in the North Adriatic Sea from March 2020 to June 2021 (see Section~\ref{sec:RelatedWorkMeasurements}). Table~\ref{tab:hydrophones} illustrates the positions and the names of the nine monitoring stations set up by the project SOUNDSCAPE. 
\begin{table}[ht]
    \centering
    \begin{minipage}[c]{0.46\textwidth}
        \includegraphics[width=\linewidth]{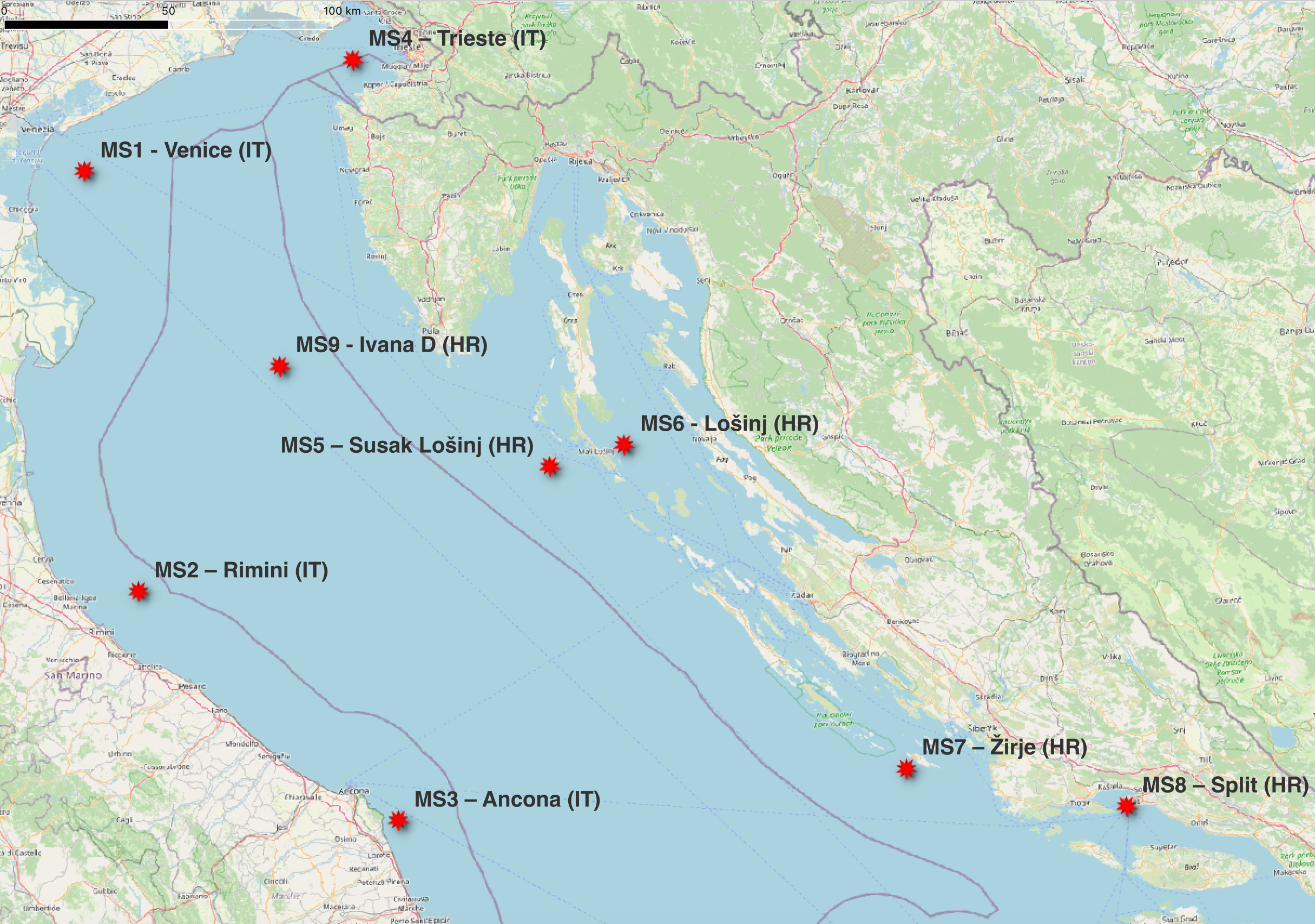}
    \end{minipage}%
    \hspace{1mm} 
    \begin{minipage}[c]{0.38\textwidth}
        \centering
        \begin{tabular}{lcc}
            \toprule
            \textbf{Monitoring station} & \textbf{Longitude} & \textbf{Latitude} \\
            \midrule
            MS1 - Venice (IT) & $12^{\circ}30.883^{\prime}$ & $45^{\circ}19.383^{\prime}$ \\
            MS2 - Rimini (IT) & $12^{\circ}42.656^{\prime}$ & $44^{\circ}10.254^{\prime}$ \\
            MS3 - Ancona (IT) & $13^{\circ}40.932^{\prime}$ & $43^{\circ}31.954^{\prime}$ \\
            MS4 - Trieste (IT) & $13^{\circ}33.917^{\prime}$ & $45^{\circ}37.095^{\prime}$ \\
            MS5 - Susak Lo\v{s}inj (HR) & $14^{\circ}17.293^{\prime}$ & $44^{\circ}29.545^{\prime}$ \\
            MS6 - Lo\v{s}inj (HR) & $14^{\circ}34.510^{\prime}$ & $44^{\circ}32.747^{\prime}$ \\
            MS7 - \v{Z}irje (HR) & $15^{\circ}36.020^{\prime}$ & $43^{\circ}37.788^{\prime}$ \\
            MS8 - Split (HR) & $16^{\circ}25.336^{\prime}$ & $43^{\circ}29.895^{\prime}$ \\
            MS9 - Ivana D (HR) & $13^{\circ}15.720^{\prime}$ & $44^{\circ}46.953^{\prime}$ \\
            \bottomrule
        \end{tabular}
    \end{minipage}
    \caption{SOUNDSCAPE hydrophones in the Northern Adriatic Sea.}
    \label{tab:hydrophones}
\end{table}

Noise levels can vary based on the frequency at which they are measured. 
The \emph{European Marine Strategy Framework Directive} (MSFD) establishes 63~Hz and 125~Hz as standard frequencies. In this work, we focus on these standard frequencies, as well as on 400~Hz and 4000~Hz, due to the increased significance of environmental factors at higher ranges. Moreover, the latter is particularly relevant for dolphins, which communicate at higher frequencies.

\subsection{Sound propagation model}
The basic objective of noise modelling is to assess how much noise a particular activity will generate in the surrounding area~\cite{farcas2016underwater}: the aim is to model the received noise level (RL) at a given point (or points), based on the sound source level (SL) of the noise source, and the amount of sound energy which is lost as the sound wave propagates from the source to the receiver (transmission loss or propagation loss, TL). The relation between these quantities is encapsulated in the classic sonar equation~\cite{urick1983principles}: 
\begin{equation}\label{eq:receiveNoiseLevelBasic}
    RL = SL - TL
\end{equation}
This straightforward expression is fundamental to modelling underwater noise, and its simplicity belies considerable complexity in the task of computing the transmission loss in order to estimate the received noise.

Sound propagation is profoundly affected by some factors such as the conditions of the surface and bottom boundaries of the sea as well as by the variation of sound speed within the ocean volume~\cite{etter2013underwater}. 
Air has a density 800 times lower than the density of water, therefore a sound that propagates inside the water has a higher propagation speed, equal to about $1500\ m/s$, against about $340\ m/s$ of air. 
So, with a sampling period of 60 seconds it makes sense to neglect propagation time within the circle of influence and, within the sampling interval, consider the noise level distribution as stationary. 
When a boat switches the engine on, we consider the noise as instantaneously propagated in the area of influence within the sampling period, without actually propagating the wavefront in space-time.

Sound propagation speed is also influenced by various chemical-physical factors such as temperature, salinity and pressure~\cite{rogers1988underwater}, varying both during the day and with the seasons in the superficial part~\cite{richardson1995marine}, and with depth.
In shallow water, that is the predominant context in Northern Adriatic sea, sound wave reflections off the seabed strongly affect propagation, and bathymetry plays an important role in determining propagation loss~\cite{ainslie2010principles}.
The computation of the transmission loss considering all these parameters is not a simple task and for this reason various models have been introduced. Before discussing the transmission loss, however, we focus on how to evaluate the source level, that is, in our case, the noise generated by the fishing vessels.

\subsection{Source level estimation}
\label{sec: sourceLevel}
The principal sources of underwater noise are machinery, propellers, and cavitation. 
Our AIS dataset includes some data of the fishing boats, such as the length overall (LOA) of the boat, the horsepower of the engine and also the fishing gear used. However, the dataset does not include  direct measurements of the sound pressure levels of the fishing vessels. So, we need to infer such values considering the general literature about underwater noise and the measurements provided by the SOUNDSCAPE project. 

A first issue is how to evaluate the increase of noise when a trawler is in action. 
Many studies focus on assessing the noise of vessels when they are free-running (i.e., when they are not performing fishing activities). However, trawling vessels typically generate higher levels of radiated noise compared to free-running vessels operating under the same machinery settings~\cite{de2006walleye}. While published data on the radiated noise from operating trawling vessels are limited, some studies have reported increases in radiated noise ranging from 5~dB to 15~dB during trawling activities~\cite{mitson1993underwater}.
Specifically, in~\cite{de2006walleye} it is noted that the effect of trawling is minimal below 100~Hz and increases with frequency. Accordingly, we assign an increase of 5~dB at 63~Hz when the vessel is trawling, 10~dB at 125~Hz, and 15~dB at higher frequencies, specifically at 400~Hz and 4000~Hz. This approach aligns with the findings in~\cite{mitson1993underwater} and reflects the change in source levels during trawling as discussed in~\cite{ghezzo2024natural}. 

To recover the sound pressure level of a specific fishing vessel, we consider a set of measurements coming from the SOUNDSCAPE project.
In particular, we use the measurements of the hydrophone MS9 located at $13^\circ15.720E$ $44^\circ46.953N$, in the middle of Adriatic Sea (see Table~\ref{tab:hydrophones}), with 42m-depth, terrigenous sandy seafloor, taken on March 31, 2021 between 5:40~pm and 5:55~pm. Here, there is a unique fishing vessel crossing nearby the hydrophone and taken as the reference boat. Thus, the recorded noise is associated to the trip 1001 of MMSI~*** (anonymized for privacy reasons) (length=27.45~m, engine power=835~Hp), while trawling at about 3.9~\rm{kn} (knots) between 500~m and 60~m from the hydrophone. 
This allows us, by linear regression on SPL measurements, to assign a vessel with an 835~Hp engine, when not trawling, an estimated source level that varies with frequency as reported in Table~\ref{table:splFrequencies}. 
\begin{table}[h]
\caption{SPL for the boat with MMSI~*** at different frequencies.}
\label{table:splFrequencies}
\begin{tabular}{ccccc}
\toprule
& {63~Hz} &125~Hz & {400~Hz} & {4000~Hz}
\tabularnewline
\midrule
SPL (1m) & 136~dB & 133~dB & 126~dB & 123~dB \tabularnewline
\bottomrule
\end{tabular}
\end{table}

In order to associate the source levels to all the other vessels, we need to relate the sound pressure level to the engine horsepower, the latter being available in our dataset. If we assume that a constant fraction of engine power gets converted into acoustic power (i.e. acoustic power scales linearly with horsepower), then 3~dB are added per doubling in engine power. We adopt such a linear progression on logarithmic scale of engine power and the resulting value is denoted with $SL_0$. For example, for engines between 100~Hp and 835~Hp, considering a frequency of 63~Hz, we obtain a range between 123~dB and 136~dB. 

Differences in source level may result from variations in speed. Specifically, as noted in~\cite{chion2019meta}, the intrinsic factor of speed can influence the broadband source level of ships according to the following relation: 
\begin{equation}\label{eq:sourceLevelSpeed}
    SL = 
    \begin{cases}
       SL_0 & \text{if} \ v \le v_0 \\
       SL_0 + 15.39~dB \times log_{10}\frac{v}{v_{0}} & \text{if} \ v > v_0 
    \end{cases}
\end{equation}
where $v_{0}=3.9~\rm{kn}$ corresponds to the speed of the reference boat and $v$ is the actual speed of the vessel. 

For simplicity, we assume that our model source boat radiates uniformly in all directions, although the acoustic signature of actual boats in navigation is louder from the side-aspect and stern-aspect than from the bow-aspect.

\subsection{Transmission loss}\label{sec:transmissionLoss}
In the ideal scenario, where surface and seabed reflections as well as absorption losses are neglected, and propagation speed is uniform, simple spherical spreading governs transmission loss~\cite{erbe2022introduction}:
\begin{align}
TL = 20 \times log_{10}(r) 
\end{align}

As observed in more realistic scenarios, sound will initially exhibit spherical spreading at short distances, where boundary effects are negligible, followed by cylindrical spreading at long ranges. 
In between, there is a transition region where neither spherical nor cylindrical spreading accurately describes the sound propagation. 
This situation can be approximated by assuming a sudden shift from spherical to cylindrical spreading at a \emph{transition range} $r_\mathit{trans}$. 
While some recommend using the water depth as a rough estimate for this transition range, it is important to use this approach with caution. As highlighted in~\cite{erbe2022introduction}, the optimal transition range $r_\mathit{trans}$ varies depending on seabed characteristics. 
Simulation by parabolic equation modelling~\cite{oliveira2021underwater} (confirmed by normal-mode modelling) shows that for a flat bottom at 50~m-depth (soft seabed, $\rho=1500~\mathrm{kg}/\mathrm{m}^{3}$, $c = 1700~\mathrm{m}/\mathrm{s}$), there is about 48~dB attenuation at 400~m, corresponding to over 8 doublings with spherical attenuation. Between 2 and 4~km the attenuation is about 4.5~dB, which is compatible with mode stripping attenuation (between spherical and cylindrical): $15\ log_{10}(r)$. 
Propagation modelling and experimental measurements in~\cite{smith2023underwater} show decreased attenuation in shallow water. For soft seabeds, propagation loss is greater than for hard seabeds, but attenuation is lower at 15~m than at 30~m depth. 
Up to 100 m, at 30 m-depth, with soft seabed, sound pressure decrease is almost identical to spherical spreading. At 15 m-depth, with soft seabed, sound pressure decrease is spherical up to about 60 m, then the decrease is much lower. 
The same study reports a 4~dB increase in sound level when speed changes from 6 to 11~\rm{kn}, confirming the velocity-dependent component of $15.39~\mathrm{dB} \times log_{10}\frac{v}{v_0}$ from Equation~\ref{eq:sourceLevelSpeed}. 
In~\cite{ainslie2010principles} Ainslie proposes a modification to cylindrical spreading for long ranges, once the reflection losses due to multiple bottom reflections begin to accumulate. Following Ainslie's model, we adjust the propagation loss to implement the combination between spherical propagation and mode stripping as follows: 
\begin{equation}\label{eq:transmissionLoss}
    TL = 
    \begin{cases}
        20\ log_{10}(r) & \text{if}\ \ r \le r_\mathit{trans}\\
        15\ log_{10}(r) + 5\ log_{10}(r_\mathit{trans}) & \text{if}\ \ r > r_\mathit{trans}
    \end{cases}
\end{equation}
The $15\ log_{10}(r)$ dependence on range is known as mode stripping because it results from the gradual erosion of steep ray paths (high-order modes) after multiple bottom reflections. 
To determine $r_\mathit{trans}$, we refer to the trajectory of trip 1001 of reference boat MMSI~***. At 63~Hz the transition is expected to occur at around 400~m, approximately 10 times the water depth. This indicates that $r_\mathit{trans}$  is parametrically dependent on depth through a multiplicative factor of 10. 
For the middle frequencies (125~Hz and 400~Hz) we set the commutation between the two propagation regimes at
4 times the depth based on our simulations. 
For the highest frequency (4000~Hz) we restrict such commutation range even further, to twice the depth.

With simple geometric spreading, the role of absorption in propagation is not accounted for. Environmental absorption features may affect the transmission loss, especially for large distances and high frequencies. %
In more realistic models, one needs to consider all the environmental aspects that influence the sound propagation underwater, by adding a term proportional to distance from the source~\cite{erbe2022introduction}:
\begin{equation}\label{eq:tlAlpha}
TL_{tot} = TL + \alpha \times r
\end{equation}
In the literature there are several models for predicting the absorption of sound in sea water which retain the essential dependence on temperature, pressure, salinity, acidity and other environmental features. 
In the Francois and Garrison model~\cite{francois1982sound1} the general equation for the absorption of sound in sea water, at a given frequency $f$, is given as the sum of contributions from boric acid, magnesium sulfate, and pure water. At a frequency below 100~Hz only the first contribution is relevant, although $\alpha$ 
is approximately of the order of $10^{-6}~\mathrm{dB}/\mathrm{m}$~\cite{erbe2022introduction}. 
Specifically, at frequencies of 63~Hz and 125~Hz, $\alpha$ is on the order of $10^{-6}~\mathrm{dB}/\mathrm{m}$, while at 400~Hz it increases to the order of $10^{-5}~\mathrm{dB}/\mathrm{m}$, and at 4000~Hz it reaches the order of $10^{-4}~\mathrm{dB}/\mathrm{m}$. 

\subsection{Ambient noise}\label{sec:underwaterSoundPropModel}
The received noise level (RL) at a given point is computed starting from Equation~\ref{eq:receiveNoiseLevelBasic}. However, the formula does not consider the ambient (or background) noise, which is present in the marine environment. 
The received noise level $RL$ exceeding ambient noise is
\begin{equation}\label{eq:receiveNoiseLevelAmbientNoise}
RL = SL-TL_{tot}-AN
\end{equation}
where $SL$ is the sound source level, $TL_{tot}$ the transmission loss and $AN$ the ambient noise. 

We use the SOUNDSCAPE measurements~\cite{picciulin2023SR,picciulin2023data} also to estimate the ambient noise. 
In particular, we employed the exceedance level $L_{90}$~\cite{vanGeel2022Abrief}, which indicates the sound level that is exceeded 90\% of the time and is equal to the $10^{th}$ percentile statistic. As mentioned in~\cite{picciulin2023SR}, $L_{90}$ can be referred to the common natural acoustic conditions.
This sound level is very different at the nine stations along the year and at the various frequencies (see Figure~3 in~\cite{picciulin2023SR}). Hence, for each frequency and for each month of the year, we compute the \emph{ambient noise map} in the following way. 
We partition the Northern Adriatic Sea into a regular grid composed of square spatial cells (1km$\times$1km).
Given a certain frequency and a month, to assign an ambient noise value to each cell of our grid, we started from the cells where hydrophones are located: we associated the exceedance level $L_{90}$ of the hydrophone at the corresponding cell. Then we applied an Inverse Distance Weighting (IDW) interpolation using QGIS\footnote{https://qgis.org/en/site/}, an Open Source GIS that supports viewing, editing, and analysis of geospatial data. In this type of interpolation, the sample points are weighted so that the influence of each point decreases with distance from the unknown point being estimated. This approach allows us to assign an ambient noise value that varies for each grid cell, capturing the differences in underwater noise across various regions of the Northern Adriatic Sea. 
For instance, in June 2020, at 63~Hz the most silent area is Ancona (Italy) with an ambient noise of 60.78~dB and the loudest zone is \v{Z}irje (Croatia) with a value of 82.62~dB. Instead, at 4000~Hz Ancona has an ambient noise of 84.65~dB and \v{Z}irje reaches a value of 92.80~dB.

\subsection{Summing different received noise levels}
At a measurement point that is equally distant from two equally-powerful sound sources, the two contributions would add up in magnitude and phase. However, distinct and independent sources, such as two boats, can be treated as incoherent sources. Even in a narrow frequency band, there will be a random phase difference between the two sources. Therefore, 
the noise in a 1/3 octave band around 63~Hz (or in any other band) gets increased by 3~dB if there are two equal contributions, by 6~dB if there are four equal contribution, etc.~\cite{Erbe2022Terminology}.
More precisely, what does add are the intensities, after inversion of the logarithmic function that defines the decibel. Generally, if we have $n$ sources reaching a
cell with $n$ different values of RL, 
the total noise level is:
\begin{equation}\label{eq:sumRL}
RL_{total} = 10 \times log_{10}(10^{RL_1/10} + \ldots + 10^{RL_n/10})
\end{equation}

\section{Model implementation using MobilityDB}\label{sec:modelmplementation}
In this section we develop a framework for the spatio-temporal characterisation of underwater noise. As already mentioned, we partition the Northern Adriatic Sea into a regular grid composed of square spatial cells (1km$\times$1km) and we estimate the noise generated by the fishing vessels in any cell at regular time intervals (every 60 seconds). 

First, as described in Section~\ref{sec: trajReconstruct}, starting from the AIS data, we create a set of semantic trajectories representing the behaviour of the fishing vessels. For this task, we build on our previous work~\cite{rovinelli2021multiple,brandoli2022multiple}. 
Then, in Section~\ref{sec: NoiseMap}, we proceed with the description of the algorithm computing the underwater noise maps. Finally, in Section~\ref{sec: MobilityDB} we provide some implementation details highlighting the advantages offered by MobilityDB~\cite{MobilityDBTODS2020}, an open source geospatial trajectory data management.

\subsection{Creation and enrichment of the trajectories of fishing vessels}
\label{sec: trajReconstruct}
The first step consists in reconstructing the trajectories of the fishing vessels starting from terrestrial Automatic Identification System (AIS) data, i.e., the AIS data sent by ships and received by ground stations on the Italian coast of Northern Adriatic Sea. AIS data contains the identifier of the vessel, called MMSI, its position and the time instant of the bearing, together with other information, like speed and course. Since boat positions are recorded every 10-20 seconds, which correspond to a small spatial displacement of the boat, trajectories are reconstructed by linear interpolation of the AIS data.  
Next, in order to organise the data into distinct trajectories followed by the fishing vessels, also called trips,  the continuous movement of a vessel  is split according to several criteria. For example, a new trip begins when the vessel is inside a port area and there is no AIS transmission for longer than a fixed time (see~\cite{brandoli2022multiple} for more details).

A trip consists of a sequence of segments obtained by connecting consecutive AIS points. The next step is to enrich such trajectories with different kinds of semantic information, called \emph{aspects}, following the MASTER model~\cite{mello2019master}. 
The model distinguishes among \emph{long-term} aspects, (associated with the full trajectory), \emph{volatile} aspects (associated with the segments) and \emph{permanent} aspects (associated with the fishing vessel, derived from the MMSI). 
A \emph{long-term} aspect is the length and the duration of the trajectory whereas a \emph{permanent} aspect, defined for this specific work, is the sound level associated with the engine horsepower of the vessel, $SL_0$. This aspect is computed as specified in Section~\ref{sec: sourceLevel}, and it is denoted by $mmsi.\mathit{SL_0}$. 
Two crucial  \emph{volatile} aspects are the \emph{speed} of the vessel and the \emph{activity} carried out by the fishing vessel. We consider the following activities: \emph{in port}, \emph{entering to} and \emph{exiting from} the port, \emph{navigation} and \emph{fishing}. The \emph{in port}, \emph{entering to} port and \emph{exiting from} port situations can be deduced from the position of the extremes of the segment w.r.t. the port area. If none of the previous cases applies, the \emph{fishing} or \emph{navigation} activities are established on the basis of the average speed of the boat. This aspect is of fundamental importance for the underwater sound propagation model, because when a boat is fishing it produces a much more intense sound. Given a spatio-temporal point $p = ((x,y),t)$ belonging to a segment $s$ ($(x,y) \in s$) in a certain time interval $I$ ($t \in I)$, we set $p.\mathit{fishing}$ to $1$ if the activity associated to the segment $s$ during $I$ is \emph{fishing}, and to $0$ otherwise. On the other hand, $p.\mathit{speed}$ denotes the speed of the vessel in $p$.

\subsection{Construction of the noise maps}
\label{sec: NoiseMap}
In this section we describe the procedure for assigning a noise level at a certain frequency $f$ induced by the fishing vessels to the cells of a regular grid, partitioning the Northern Adriatic Sea, every 60 seconds. We consider a set of spatio-temporal cells $\mathbb{G}= \mathbb{S} \times \mathbb{T}$ where $\mathbb{S}$ is a regular grid consisting of 1km$\times$1km spatial cells, and $\mathbb{T}$ is a set of time instants, such that $t_0$ is a fixed time instant and $t_{i+1} = t_i + 60$. 
Hence, each spatio-temporal cell $c \in \mathbb{G}$ consists of two components, $(g, t)$, representing the spatial cell $g$ at time instant $t$. 
For each frequency $f \in \{63, 125, 400, 4000\}$ we write $\mathbb{G}_f$ to denote the set of spatio-temporal cells annotated with pieces of information which possibly depend on the chosen frequency. The annotations are: (i) \emph{ctd} contains the coordinates of the centroid of $g$; (ii) \emph{depth} stores the depth of the sea in $c$; (iii)
$\alpha$ stores the absorption of sound as defined in Section~\ref{sec:transmissionLoss}; (iv) \emph{an} records the ambient noise in $c$ estimated as reported in Section~\ref{sec:underwaterSoundPropModel}; (v) \emph{rl} records the total noise received in $c$, i.e., by the centroid of $g$ at time instant $t$. It is worth noticing that only the latter three annotations depend on the frequency \emph{f}.

Let $\cal TR$ be the set of the trajectories of the fishing vessels, $\mathbb{G}_f$ be the spatio-temporal grid with \emph{f} $\in \{63, 125, 400, 4000\}$. 
Algorithm~\ref{alg:RL} computes the total received noise level for every cell $c \in \mathbb{G}_f$ at the frequency \emph{f}. We use $p\! \downarrow \! 1$ to denote the projection on the first component of the spatio-temporal point $p$, i.e., its  coordinates, and \text{d}($z_1$, $z_2$) for the Euclidean distance between two spatial points $z_1$ and $z_2$.

\begin{algorithm}
\caption{Given $\cal TR$, $\mathbb{T}$, $\mathbb{G}_f$ and \emph{f}\, $\in \{63, 125,400, 4000\}$, the algorithm computes the total received noise level for each $c \in \mathbb{G}_f$ at frequency \emph{f}.}
\label{alg:RL}
\small{\begin{algorithmic}[1]
\State Let $mp$: $\mathit{map}\langle \mathit{cell}, \mathit{float} \rangle$
\For{each $tr \in \cal{TR}$}
    \For{each $t \in \mathbb{T}$}
        \State p = $(tr(t), t)$
        \label{line: point}
        \State $c_p$ = unique cell $c\in G_f$ such that $p \in c$
        \label{line: cell}
        \State $v_0 = 3.9$
        \If {$p.\mathit{speed}>v_0$}
        \label{line:SLstart}
            \State $SL = tr.\mathit{mmsi}.SL_0 + 15.39 \cdot log_{10}\frac{p.\mathit{speed}}{v_0} +\mathit{inc}_f \cdot p.\mathit{fishing}$ 
        \Else 
            \State $SL = tr.\mathit{mmsi}.SL_0  +\mathit{inc}_f \cdot p.\mathit{fishing}$ 
        \EndIf \label{line:SLend}
        \State $r_\mathit{trans} = c_p.\textit{depth}\cdot \mathit{mult}_f$ 
        \label{line: trans}
        \State \label{line: r}  $r = 10^{(SL - 5\cdot \log_{10}(r_{\mathit{trans}}) - c_p.\mathit{an})/15}$ 
        \For{each $c = (g, t)\in \mathbb{G}_f . \ \ 
        \text{d}(c.\mathit{ctd},p\!\downarrow\!1) < r$} \label{line: propS}
            \State $\mathit{dist} = \textrm{d}(c.\mathit{ctd}, p\!\downarrow \!1)$
            \If {$\mathit{dist} \leq r_{\mathit{trans}}$} 
                \State $RL = SL - 20 \cdot \log_{10}(dist) - c.\alpha \cdot dist - c.an$
            \Else 
                \State $RL = SL - 15 \cdot \log_{10}(\mathit{dist}) - 5 \cdot \log_{10}(r_{\mathit{trans}}) - c.\alpha \cdot \mathit{dist} - c.an$
            \EndIf
            \State $\mathit{mp}[c] = \mathit{mp}[c] + 10^{RL/10}$ \label{line: propE}
        \EndFor
    \EndFor
\EndFor

\For{each $c \in \mathbb{G}_f$}
    \State $c.rl = 10 \cdot \log_{10} (mp[c])$
    \label{line:RL}
\EndFor
\end{algorithmic}}
\end{algorithm}

The noise is estimated every 60 seconds at the selected frequency, i.e., in the time instants belonging to $\mathbb{T}$ at frequency $f$. The centroids of the grid cells are considered as listening points (we have $43,386$ of these points), and consequently the noise \emph{received} at a centroid point models the noise in all the points of the cell at a certain time instant and at frequency $f$.

In order to build the noise map at frequency $f$, we get the positions of all the fishing vessels at the same time instants, i.e., every 60
seconds. For each point (Line~\ref{line: point}), we determine the cell it belongs to (Line~\ref{line: cell}) and we calculate the noise generated by the fishing vessel (Lines~\ref{line:SLstart}--\ref{line:SLend}) obtained by adding to the sound level associated with the horsepower of the boat (\emph{mmsi.SL}$_0$), a contribution related to the actual speed of the vessel in $p$ 
(see Equation~\ref{eq:sourceLevelSpeed}), 
and the noise due to the fishing activity if it occurs in $p$. Notice that the latter, \emph{inc}$_f$, can range from 5~dB up to 15~dB depending on the frequency as discussed in Section~\ref{sec: sourceLevel}. In Line~\ref{line: r}, the sound propagation radius $r$ (expressed in meters), i.e., the distance at which the noise generated by the fishing vessel gets drowned into ambient noise, is computed. This is obtained by using Equation~\ref{eq:receiveNoiseLevelAmbientNoise} and setting the received noise ($RL$) to $0$:
$$0 = SL-TL_{tot}-AN$$
In computing the radius we ignore the coefficient of absorption $\alpha$ in Equation~\ref{eq:tlAlpha} for $TL_{tot}$, allowing for a simplification of calculation and getting an overestimation of $r$, hence the approximation is safe. With some simple mathematical steps we get $r = 10^{(SL - 5\cdot \log_{10}(r_{\mathit{trans}})-c_p.\mathit{an})/15}$ where $r_{\mathit{trans}}$ is the transition range when spherical propagation shifts to a lower attenuation. Its value depends on the sea depth in $c_p$ (Line~\ref{line: trans}) and it varies according to frequency $f$, $\mathit{mult}_f$, ranging from 2 up to 10 times the sea depth, as explained in Section~\ref{sec:transmissionLoss}.
Then, we propagate the noise in the cells that are within the radius $r$ (Lines~\ref{line: propS}--\ref{line: propE}) according to Equation~\ref{eq:transmissionLoss} and we compute the relative received noise level by Equation~\ref{eq:receiveNoiseLevelAmbientNoise}. Finally, by using Equation~\ref{eq:sumRL}, we combine all the received sound levels to obtain the total noise level at frequency $f$ to be associated with the cell (Line~\ref{line:RL}).  

Concerning the complexity, let $n = |\cal TR|$, $m = |\mathbb{T}|$, $k = |\mathbb{G}|$, $a$ be the area of a grid cell and $r$ the largest radius arising in Line~\ref{line: r}. Then the complexity is $O(n \cdot m \cdot r^2/a + k)$. The factor $r^2/a$ is motivated by the fact that in Line~\ref{line: propS} we consider the cells in a neighbourhood of radius $r$. Note that $r$ depends on the source level, which is bounded by the maximum engine power and the maximum speed of the monitored fishing vessels. 
In our case, the radius $r$ for each frequency is less than 71,254~m for 63~Hz, 111,399~m for 125~Hz, 31,802~m for 400~Hz, and 5,478~m for 4000~Hz. 

In order to process all this data and build our model, we used a machine that features 32 Intel(R) Xeon(R) CPU E5-4610 v2 processors running at 2.30 GHz, offering multithread performance. It is equipped with 256 GB of DDR4 ECC RAM and it utilises a 500 GB RAID 5 storage configuration. 
We evaluated the time for constructing the model assuming a daily data processing and considering the month of June 2020. Our model requires approximately 20 minutes per day for the construction and propagation of underwater noise at 63~Hz, 22 minutes at 125~Hz, 10 minutes at 400~Hz, and 4 minutes at 4000~Hz. 

\subsection{Implementation details}
\label{sec: MobilityDB}
To construct and store the set of trajectories and to implement the underwater noise model, we used MobilityDB~\cite{MobilityDBTODS2020}, a moving-object database that extends the type system of PostgreSQL and PostGIS with abstract data types supporting temporal types and spatio-temporal operators to manage moving objects. 
The offered constructs perfectly suit the representation of trajectories, which can be  reconstructed from a sequence of spatio-temporal data, and allow for semantic enrichment of trajectories. Moreover, it offers spatial and spatio-temporal indices to improve the efficiency of the general procedure described in Algorithm~\ref{alg:RL}.

Let us present the structure of the table storing the trips of the fishing vessels: 
\begin{tabbing}
\qquad \= CREA \= \kill
\> \texttt{CREATE TABLE vessel\_trip (} \\
\> \>    \texttt{index integer PRIMARY KEY GENERATED ALWAYS AS IDENTITY,} \\
\> \>     \texttt{trip\_id integer,} \\
\> \>     \texttt{mmsi integer NOT NULL,} \\
\> \>     \texttt{vessel\_name character varying,} \\
\> \>     \texttt{speed tfloat,} \\
\> \>     \texttt{activity tint,} \\
\> \>     \texttt{trip tgeompoint,} \\
\> \>     \texttt{traj geometry} \\
    \> \texttt{);}
\end{tabbing}
Each trip is identified by the attribute \texttt{trip\_id} and it is related to the fishing vessel having \texttt{mmsi} and \texttt{vessel\_name}. The attributes \texttt{speed} and \texttt{activity} model the \emph{volatile} aspects representing the speed and the activity of the vessel during the voyage and the attribute \texttt{trip} models the spatial coordinates of the movement followed by the vessel. These three attributes have temporal types, allowing for the representation of the variation in time of the speed, the activity and the position of the fishing vessel starting from the AIS data. The values between successive instants are interpolated using a linear function for \texttt{speed} and \texttt{trip} whereas for \texttt{activity} a step function is used. 
Finally, the attribute \texttt{traj} with type \texttt{geometry} is employed to visualise the trajectory.

To improve the spatial operations on trajectories, like \texttt{ST\_Intersects}, we add a spatial index on the attribute \texttt{traj} of the table \texttt{vessel\_trip}:
\begin{verbatim}
    CREATE INDEX Vessel_Trip_Geom_Idx ON vessel_trip USING gist(traj);
\end{verbatim}
Indexing speeds up searching by organising the data into a search tree which can be quickly traversed to find a particular record. The spatial index structure used is \emph{R-Tree}.

\begin{figure}[ht]
  \centering
\includegraphics[width=\linewidth]{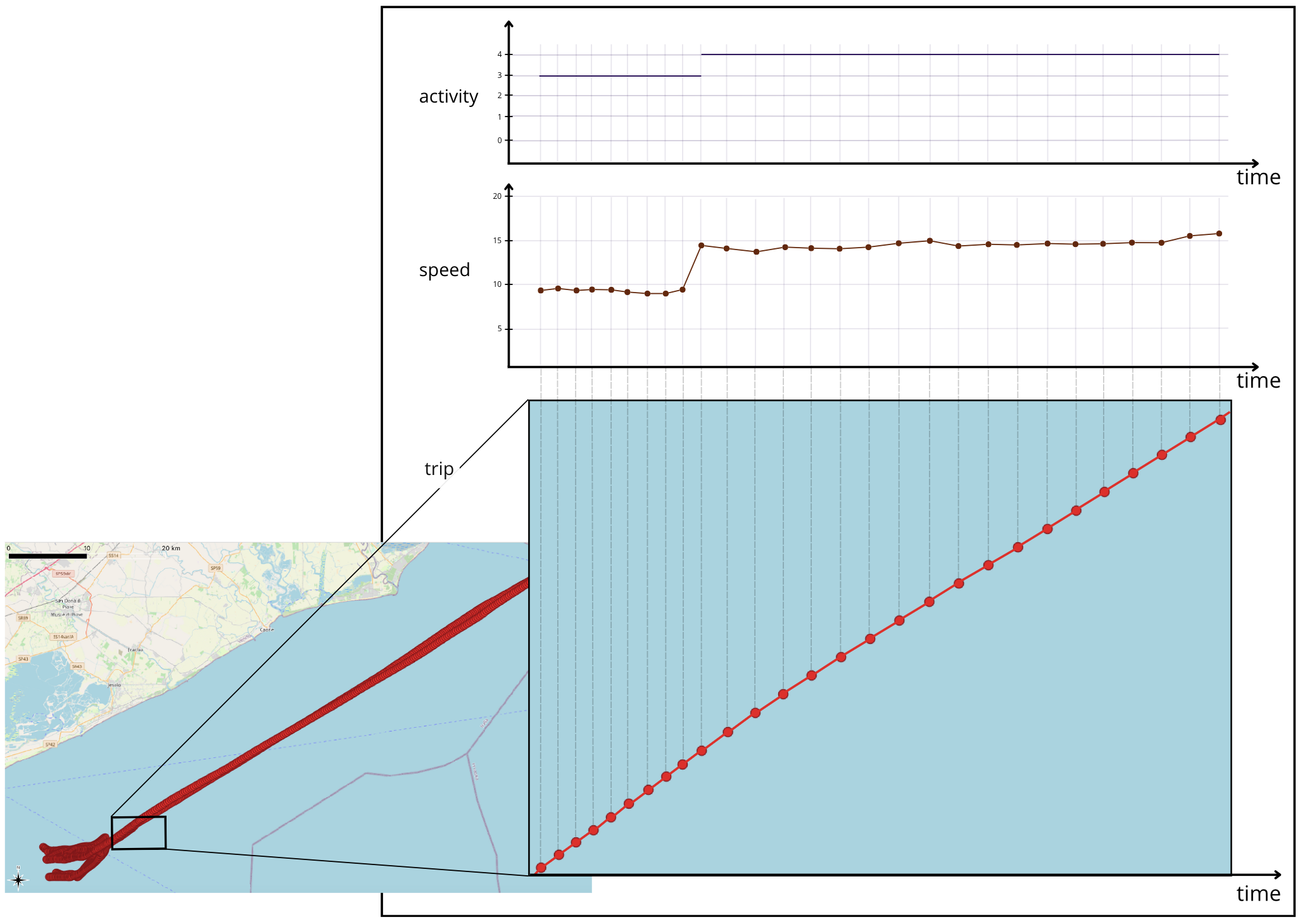}
\caption{Representation of the three temporal types \texttt{activity}, \texttt{speed} and \texttt{trip}.}
  \label{fig:trajectoryMobilityDB}
\end{figure}

Once reconstructed the trajectories from the AIS data, we need to get the values of all the trajectories at the same time instants, every $60$ seconds.
MobilityDB offers an efficient function, \texttt{tsample()} to sample a temporal value according to period buckets. We apply such a function to get speed, activity and position every minute, i.e.,
\texttt{tsample(speed,'1 min')}, \texttt{tsample(activity,'1 min')}, \texttt{tsample(trip,'1 min')} and these values are inserted into the table \texttt{vessel\_trip} for the attributes \texttt{speed}, \texttt{activity} and \texttt{trip}, respectively. 
In this way the three temporal values are built on the same set of minutes which is $\mathbb{T}$. Figure~\ref{fig:trajectoryMobilityDB} displays an example of the values for the attributes \texttt{activity}, \texttt{speed} and \texttt{trip}. The vertical lines represent the time instants, the three attributes are synchronised and the sample period is one minute. \texttt{Trip} represents the movement of the fishing vessel, which consists of a sequence of points whose distance between consecutive points becomes larger because the speed increases. In fact the temporal value \texttt{speed} shows a variation from 10~\rm{kn} up to 15~\rm{kn}. This change of speed causes a shift in activity as well: the attribute \texttt{activity} varies from $3$ denoting \emph{fishing} to value $4$ indicating  \emph{navigation}.

\begin{figure}[ht]
  \centering
\includegraphics[width=\linewidth]{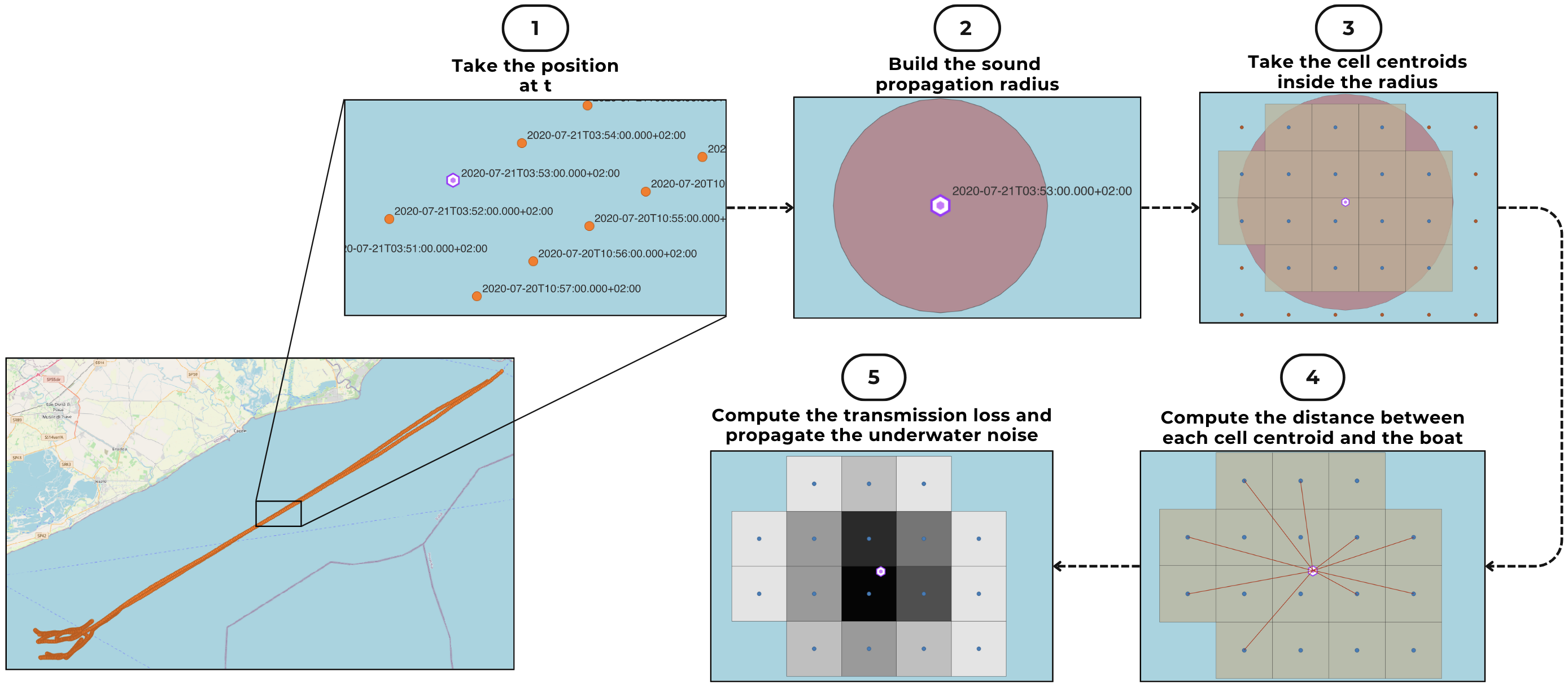}
\caption{Main steps in the calculation of the noise maps.}
  \label{fig:stepbystep}
\end{figure}

In order to compute the total received noise level for each cell of our grid at a given frequency $f$, we proceed as specified by Algorithm~\ref{alg:RL} and illustrated in Figure~\ref{fig:stepbystep}. For each spatio-temporal point $p$ belonging to \texttt{trip} we compute the propagation radius $r$. By using the function \texttt{ST\_Buffer} we build, around the spatial coordinates of $p$, a buffer $b$ with radius $r$ (Step 2 in Figure~\ref{fig:stepbystep}). Then, we select all the cells whose centroids are inside $b$ through the predicate \texttt{ST\_Intersects} (Step 3 in Figure~\ref{fig:stepbystep}) and we compute the distance between the point $p$ and these centroids (Step 4 in Figure~\ref{fig:stepbystep}). We use this distance to estimate the transmission loss which allows us to determine the received noise level in the selected cells. By grouping by cell id and time, we combine all the contributions of the points of the different trajectories through Equation~\ref{eq:sumRL} (Step 5 in Figure~\ref{fig:stepbystep}).

\section{Analyses and results}
\label{sec: results}
We next present some analyses performed by using our spatio-temporal characterisation of the underwater noise (Section~\ref{sec: experiments}). 
In this study, we analysed the complete AIS dataset for fishing vessels throughout the year 2020, comprising $92,916,965$ AIS records from $714$ unique vessels. It is worth recalling that 2020 is a period of restricted shipping activity due to COVID-19 pandemic. In Italy from March 9, 2020 until May 18, 2020 a lockdown was imposed, during which many activities, including restaurants, had to close or limit their business. 
Other containment measures were adopted also in Autumn until the end of the year. 

Moreover, in Section~\ref{sec: validation} we outline some considerations on model validation, discussing how our model can be compared with the spatio-temporal characteriziation provided by the SOUNDSCAPE project. 

\subsection{Some experiments}
\label{sec: experiments}
The first experiment consists of assessing the impact of the fishing activity on the underwater noise level considered at different frequencies. Our framework allows us to perform this analysis according to different time granularities, e.g., yearly, seasonally, monthly or daily. We chose to focus on June 2020 because it is one of the months with the highest fishing activity in the year 2020. 
Table~\ref{table:datasetSound} reports the number of vessels, AIS data and trips in the whole year 2020, in April and June 2020.  
\begin{table}[ht]
\caption{No. of vessels, AIS data and trips for year 2020, April, and June 2020.}
\label{table:datasetSound}
\begin{tabular}{lccc}
\toprule
\textbf{Period}& \textbf{Vessels}&\textbf{AIS data}  & \textbf{Trips}
\tabularnewline
\midrule
Year 2020 & 714 & $92,916,965$ & $72,776$ \tabularnewline
April 2020 & 548 & $5,392,677$  & $4,923$ \tabularnewline
June 2020 & 642 & $9,841,079$  & $7,462$ \tabularnewline
\bottomrule
\end{tabular}
\end{table}

As explained in Section~\ref{sec:UnderwaterModel}, we consider the frequencies $63~\rm{Hz}$, $125~\rm{Hz}$, $400~\rm{Hz}$, and $4000~\rm{Hz}$. 
For each frequency, we created a bivariate map for June, illustrating two variables at once. 
The first variable represents the average underwater noise that exceeds the background noise, while the second indicates the percentage of days in the period of analysis (1 month) in which a cell is active. For a cell $c$, a day $d$ is called \emph{active} if the received sound level on $d$ in $c$ exceeds the background noise. 
Bivariate maps are crucial for understanding underwater pollution, as the damage on living species depends both on noise level and noise persistence. 
Elevated noise levels over a short duration do not affect marine life as significantly as noise at prolonged intervals, which can have more severe consequences for aquatic ecosystems. 

\begin{figure*}[ht]
    \centering
    \begin{subfigure}{0.46\textwidth}
    \includegraphics[width=\linewidth]{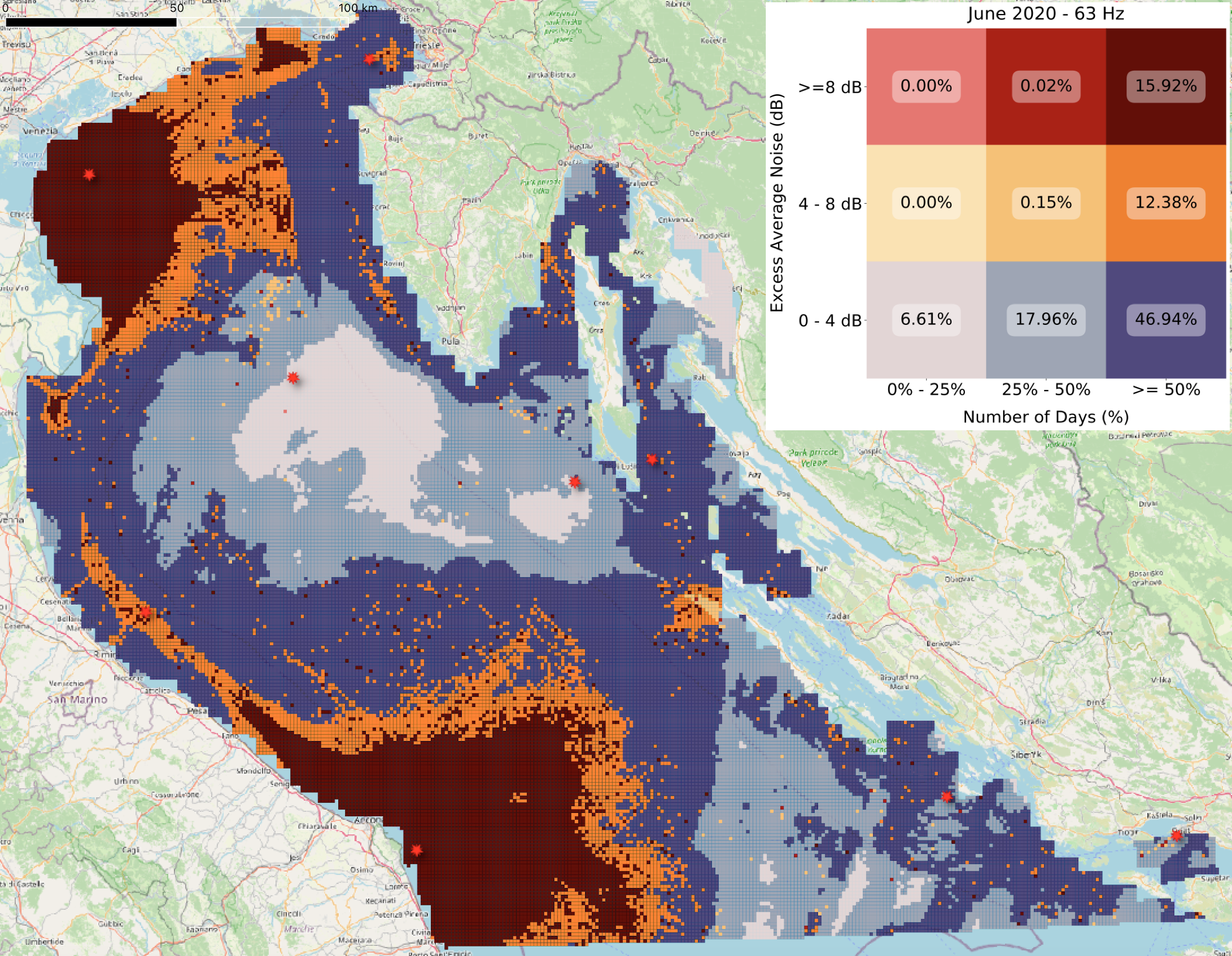}
    \caption{63~Hz.} \label{fig:june2020_10perc_63hz}
    \end{subfigure}
    \hspace{1mm}
    \begin{subfigure}{0.46\textwidth}
    \includegraphics[width=\linewidth]{./images/06_2020/june2020_10perc_125hz}
    \caption{125~Hz.} \label{fig:june_10perc_125hz}
    \end{subfigure}
    \hspace{1mm}
    \begin{subfigure}{0.46\textwidth}
    \includegraphics[width=\linewidth]{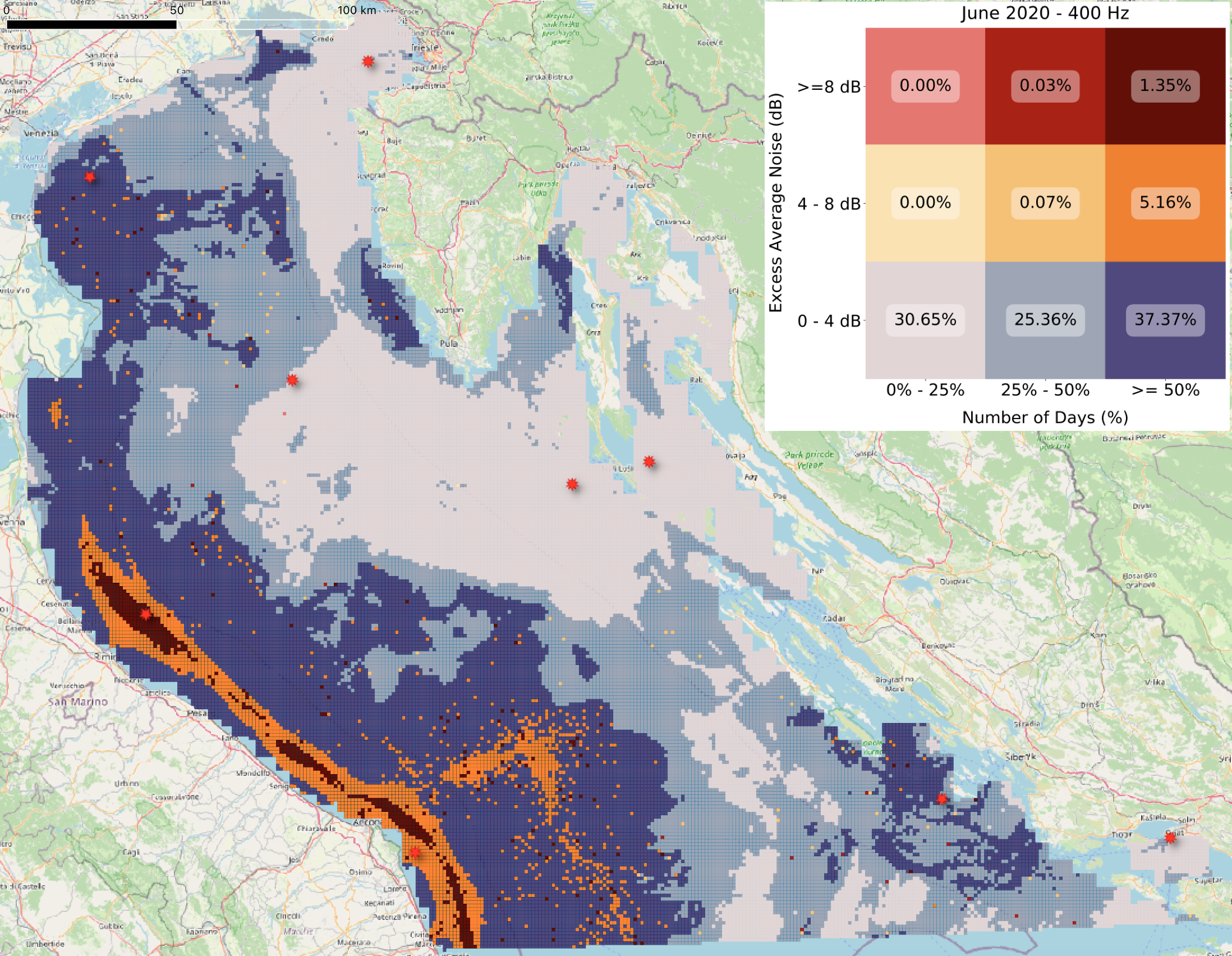}
    \caption{400~Hz.} \label{fig:june2020_10perc_400hz}
    \end{subfigure}
    \hspace{1mm}
    \begin{subfigure}{0.46\textwidth}
    \includegraphics[width=\linewidth]{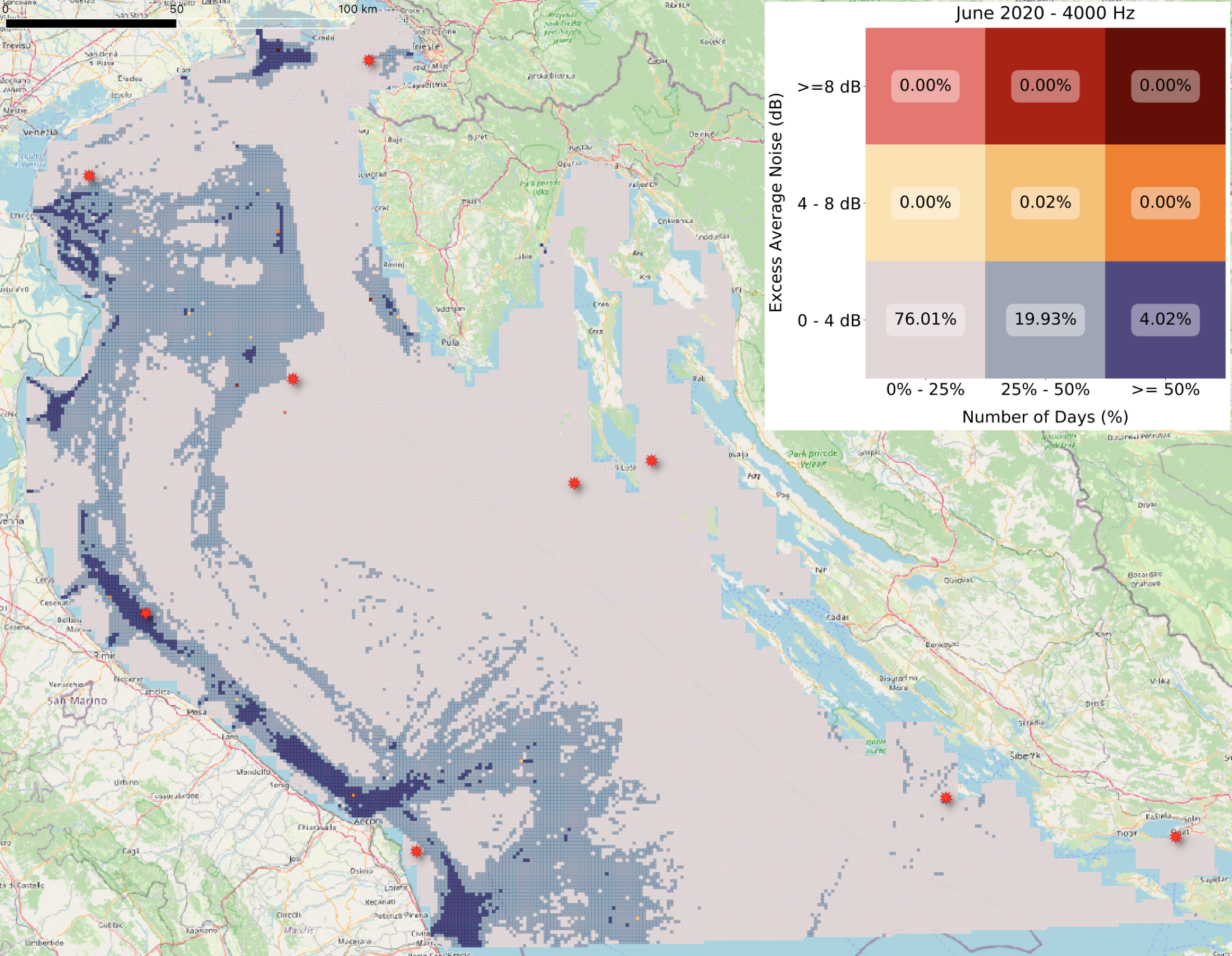}
    \caption{4000~Hz.} \label{fig:june2020_10perc_4000hz}
    \end{subfigure}
    \caption{Underwater noise bivariate maps for June 2020. The red stars are the hydrophones of SOUNDSCAPE.}
    \label{fig:june_10perc_BivariateMaps}
\end{figure*}
In Figure~\ref{fig:june_10perc_BivariateMaps}, we can observe the bivariate maps for June for the frequencies of 63~Hz, 125~Hz, 400~Hz, and 4000~Hz. 
This figure clearly demonstrates that the impact of fishing vessels varies significantly depending on the frequency analysed. These variations stem from several frequency-dependent factors: the initial SPL ($SL_0$), the absorption coefficient ($\alpha$), the increased noise generated by the vessel during fishing activities, and the background noise, which rises with higher frequencies. For example, at 63~Hz, background noise starts at 60.78~dB, whereas the lowest value at 4000~Hz is 81.07~dB. 
Figure~\ref{fig:june2020_10perc_4000hz} shows that at 4000~Hz, fishing vessels have a negligible impact on the underwater environment: all the cells are characterised by excess noise levels below 4~dB and only 4\% of the study area exhibits persistence greater than 50\%, i.e., two weeks (dark blue cells). In the majority of the cells, 76\%, the excess noise has a low persistence, below one week (light pink cells). Figure~\ref{fig:june2020_10perc_400hz} displays that at 400~Hz still 93\% of the basin presents excess noise level below 4~dB but now there is an area along the Italian coast, including several harbors, like Rimini and Ancona, extending southward to Civitanova Marche (which marks the southernmost boundary of the map) where the excess noise level is between 4~dB and 8~dB for a persistent period (more than $50\%$ of the days - dark orange cells) reaching the peak over 8~dB in a very limited zone (only 1.35\%). Besides, the area with low excess noise level and low persistence, depicted in light pink, is less than half (around 31\%) of that obtained for the frequency 4000~Hz.

Figure~\ref{fig:june2020_10perc_63hz} and Figure~\ref{fig:june_10perc_125hz} illustrate that at 63~Hz and 125~Hz, fishing vessels have a significant impact on the ecosystem. In fact, more than 20\% of the basin, in areas of the Northern Adriatic Sea that are most frequently used for fishing, such as in front of Venice and near Ancona and along the Italian coast, the excess noise levels are over 4~dB with large persistence (more than two weeks -- dark orange and dark red cells). At 63~Hz there is the highest percentage of cells (almost 16\%) with the highest excess noise levels (greater than 8~dB) for a long time (over two weeks -- dark red cells) and the lowest percentage (6.6\%) of cells with excess noise levels under 4~dB for less than 1 week (light pink cells).

\begin{figure*}[ht]
\centering
\begin{subfigure}{0.47\textwidth}
\includegraphics[width=\linewidth]{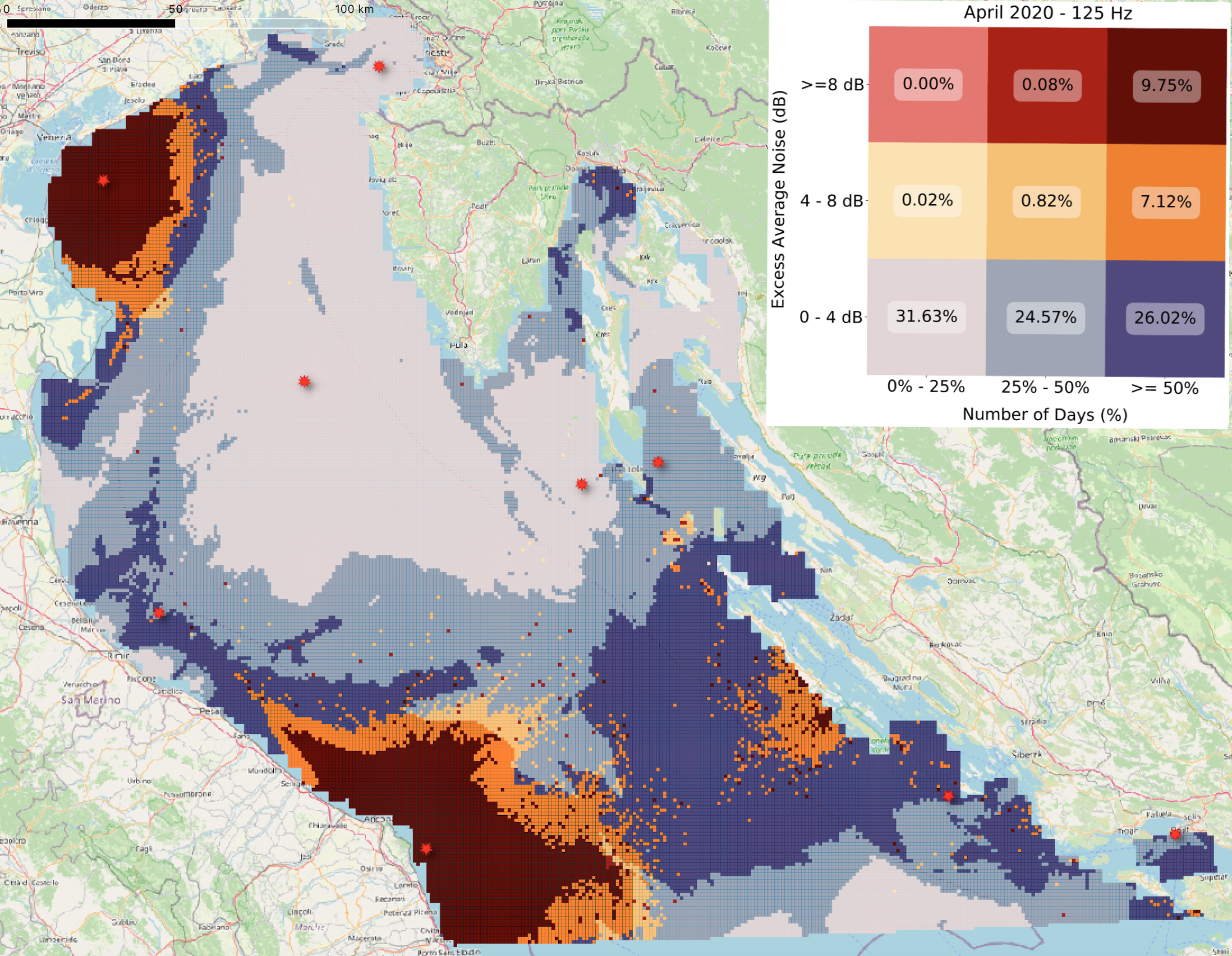}
\caption{April.} \label{fig:apr10perc125hzbivariate}
\end{subfigure}
\hspace{1mm}
\begin{subfigure}{0.47\textwidth}
\includegraphics[width=\linewidth]{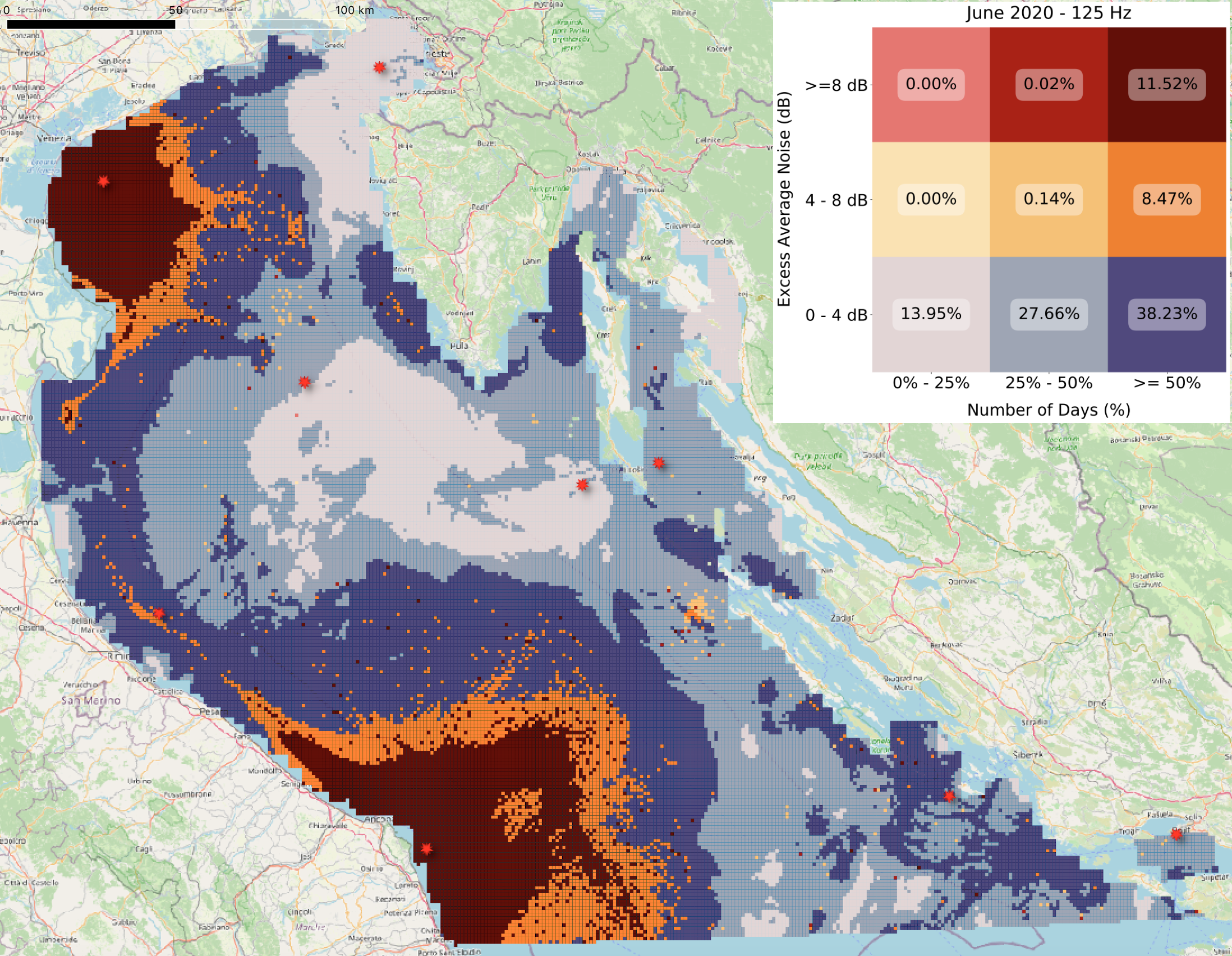}
\caption{June.} \label{fig:june10perc125hzbivariate}
\end{subfigure}
\caption{During, and post-Covid underwater noise at 125~Hz for the months of April, and June 2020.}
\label{fig:preDuringPostCovid10perc}
\end{figure*}

In order to investigate the effect of the COVID-19 pandemic outbreak on underwater sound pressure levels due to fishing activities in the Northern Adriatic Sea, we focus on April and June 2020. 
While April 2020 is a month during the lockdown, June 2020 represents the post-lockdown period, during which fishing activities gradually returned to pre-pandemic levels, reflecting the easing of restrictions and a progressive resumption of usual activities. Looking at Table~\ref{table:datasetSound}, we observe that the number of AIS data in April is limited, despite the fact that during this month usually there is an intense fishing activity.  
To compare the during-, and post-lockdown periods, we generated two bivariate maps representing the average underwater noise at 125~Hz that exceeds the background noise, along with the persistence of this noise (i.e., the percentage of active days for each cell w.r.t. the total days of the month). 
Figure~\ref{fig:apr10perc125hzbivariate} illustrates the excess noise map for April, during the lockdown. There is a large portion of the basin, i.e., 82\%, with an excess noise level below 4~dB and for 31\% of the study area there is also a low persistence (below one week -- light pink cells), thus resulting in a very silent sea. The noisest areas, with excess noise level over 8~dB for a persistent period (more than half a month -- dark red cells), are located in the Venice zone, including Chioggia, and from Ancona to the southernmost part of the map. There is also a zone in front of the Croatian coast which shows an increase of noise (excess level between 4~dB and 8~dB and even more 8~dB in several cells) for a persistent period (more than half a month). It is noteworthy that the cells with higher noise levels are located along the coast, and this confirms a redistribution of the fishing grounds, being mainly located near the coasts and in the proximity of the origin harbours as documented in~\cite{russo2021lockdown}. This behaviour could be due to the possibility to reduce time at sea, limiting the fuel consumption and the related costs.   
In June (Figure~\ref{fig:june10perc125hzbivariate}), there is a significant increase in fishing activities w.r.t. April, as witnessed by the number of AIS data and trips. In the two mentioned zones, near Venice and near Ancona, we observe a boost in the number of cells characterised by elevated noise levels and high persistence (around 3\%, dark orange and dark red). Notably, vessels ventured further offshore towards Croatia. Additionally, increased activity is evident in the region near Rimini, indicating greater exploitation of the area. 
In fact, in June only 14\% of the basin is characterised by low noise levels and low persistence (light pink cells). Conversely, cells with noise levels below 4~dB but with persistence greater than $50\%$ (dark blue cells) increase by $12\%$. 
These maps clearly points out that the limited fishing activity during the lockdown has caused a reduction in underwater noise, thus proving the acoustic impact of fishery in the Northern Adriatic Sea.

\begin{figure*}[ht]
\centering
\includegraphics[width=0.6\linewidth]{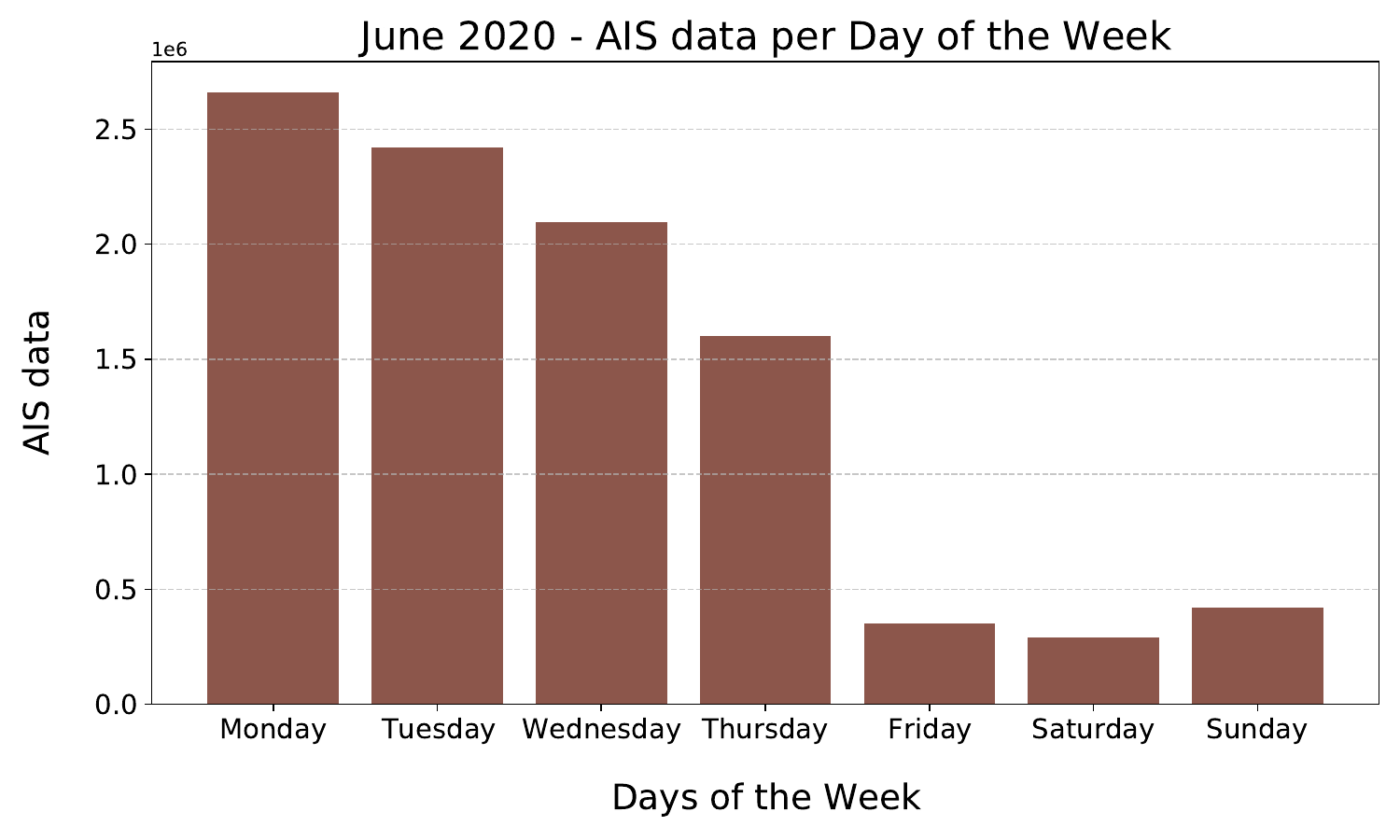} 
\caption{Number of AIS data per day of the week in June 2020.}
\label{fig:aisDataJune}
\end{figure*}

\begin{figure*}[ht]
\centering
\begin{subfigure}{0.49\textwidth}
    \includegraphics[width=\linewidth]{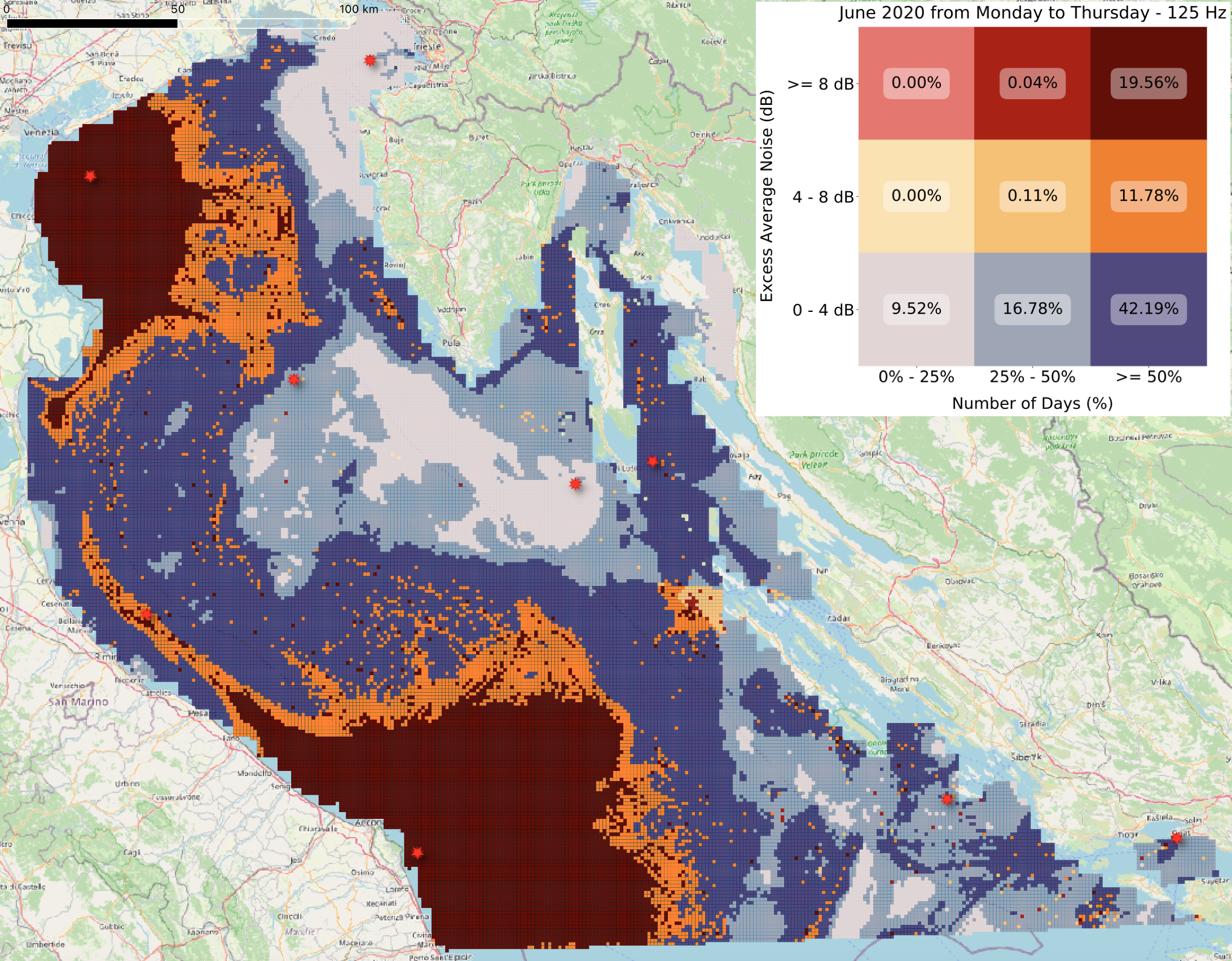}
    \caption{Average underwater noise.} 
    \label{fig:125hz_10perc_lun_gio_avg}
\end{subfigure}
\hspace{1mm}
\begin{subfigure}{0.49\textwidth}
    \includegraphics[width=\linewidth]{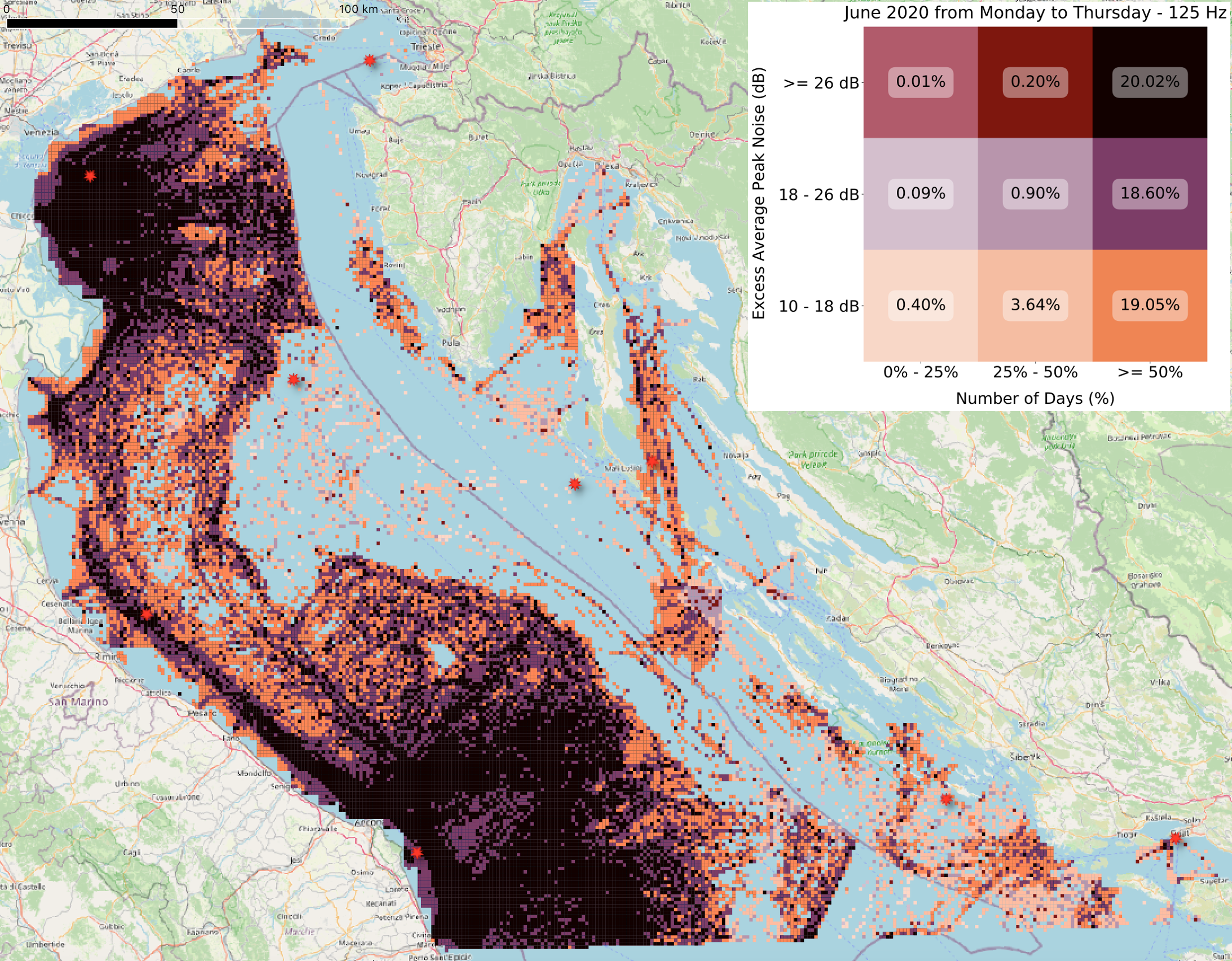}
    \caption{Underwater noise peaks. 
    } 
    \label{fig:125hz_10perc_lun_gio_peak}
\end{subfigure}
\caption{Underwater noise in June 2020 at a frequency of 125~Hz from Monday to Thursday.}
\label{fig:125hz_lun_gio_10perc}
\end{figure*}

In order to deepen the analysis on the areas with the highest noise pressure in the month of June, we investigate how the fishing activity is performed along the days of the week. 
Figure~\ref{fig:aisDataJune} illustrates the number of AIS data in June 2020 grouped by the day of the week. The plot clearly indicates that fishing vessels concentrate their activity between Monday to Thursday whereas in the weekend a very limited number of vessels go fishing. 
Hence we restrict our analysis on the days from Monday to Thursday. This is a great advantage offered by our tool, the possibility to analyse data at different granularity levels. 
Figure~\ref{fig:125hz_10perc_lun_gio_avg} presents the bivariate map illustrating the average noise that exceeds the background noise, along with the percentage of days during the week (from Monday to Thursday) when the cells are active. This analysis allows us to highlight the areas swept by fishing vessels during the four days of their highest fishing activity. Notably, we observe an increase in excess noise levels generated during these days and a reduction of the quiet cells. Compared to Figure~\ref{fig:june10perc125hzbivariate}, which considers all days of the week, there is a $15\%$ increase in areas where persistence exceeds $50\%$: the extension of the area with excess noise levels over 8~dB (dark red) is augmented of $8\%$ while the other two areas, i.e, between 4~dB and 8~dB (dark orange) and below 4~dB (dark blue) are enlarged by about 3.5\%.  
On the other hand, the number of cells with excess noise levels below 4~dB and for less than 25\% of days decreases by 4.4\%. Clearly, the two regions most exploited by fishing vessels - the area in front of Venice and the entire area around and in front of Ancona - remain the same, although with greater noise intensity as well as increased persistence. 
To further investigate the noise generated by vessels on these days, we focused on the noise peaks associated with each cell. In particular, Figure~\ref{fig:125hz_10perc_lun_gio_peak} presents a bivariate map where
the first variable represents the mean of daily peaks in June 2020 (on a scale starting from 10~dB), while the second variable represents the percentage of active days throughout the month, restricted to Monday through Thursday. 
Considering the average peaks for the month, the resulting noise is significantly higher; in fact, the bivariate map represents only values exceeding 10~dB. We observe that $19\%$ of the basin includes values between 10~dB and 18~dB, $18.6\%$ 
are characterised by underwater noise levels between 18~dB and 26~dB, and finally, $20\%$ exhibit noise levels greater or equal than 26~dB, all characterised by a high persistence. 
Only $37\%$ of the study area has a peak less than 10~dB. 
This figure clearly points out that the area in proximity of a harbour is nosier and in particular the nosiest zones are located at the south of Venice, in front of Chioggia, as well as the entire zone of Ancona and southward close to Civitanova Marche.  Moreover, this map allows us to detect the main routes of the fishing activities.   

\begin{figure}[ht]
  \centering
  \includegraphics[width=\linewidth]{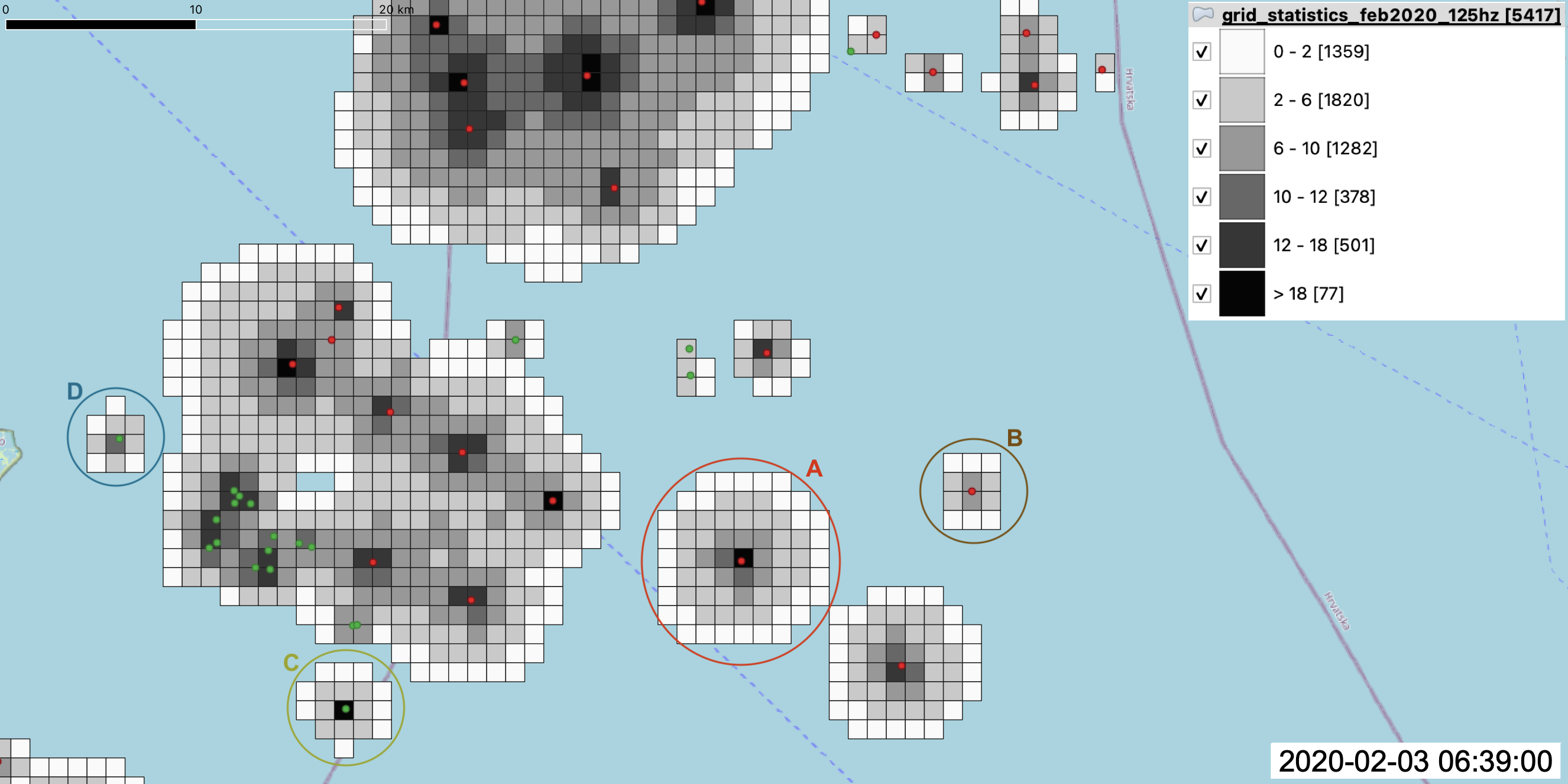}
\caption{Propagation of underwater noise of fishing vessels at 125~Hz in the Northern Adriatic Sea on February 3, 2020, at 06:39. Vessels marked with a red dot are fishing, while those marked with a green dot are not fishing.}
  \label{fig:underwaterNoise}
\end{figure}

Finally, our implementation provides also the possibility of visualising the spreading of underwater noise in time for a set of vessels. By using \emph{QGIS TimeManager}, it is possible to generate animations which, for a selection of vessels, visualise the noise propagation determined by these vessels moving in the Northern Adriatic Sea. The user can choose the boats according to several criteria, such as the range of horsepower, the MMSI, the length overall, or the activity, and the time window of the analysis. In Figure~\ref{fig:underwaterNoise} we can observe a different sound propagation at 125~Hz depending on the engine power of the vessel, its speed and its activity. 
We focus on four vessels: vessel $A$ has engine power 590.9~Hp, vessel $B$ 613~Hp and vessel $C$ 649.9~Hp (all three with the same $SL_0$ 133~dB), while vessel $D$ has engine power 215.7~Hp ($SL_0$ 130~dB). We can observe that both vessel $A$ and $B$ are fishing (red dot), but vessel $A$ is moving at a speed of 5.67~\rm{kn}, while vessel $B$ is moving at 2.5~\rm{kn}. It is worth to remark that the sound propagates across more cells (covering more kilometers) for vessel $A$ compared to vessel $B$, illustrating how speed affects sound propagation. Vessel $C$ has the same source level as vessels $A$ and $B$ but is not fishing (green dot) and is moving at a speed of 10.75~\rm{kn}. Compared to vessel $A$, despite vessel $C$ is moving at a higher speed, the sound propagates less underwater. Specifically, sound propagates approximately 3~km for vessel $C$ and 5 km for vessel $A$, highlighting the greater impact of fishing activity compared to the vessel's speed. Finally, vessel $D$ has the same speed as vessel $C$ (10.48~\rm{kn}), and it is not fishing (green dot), but its $SL_0$ is 3~dB lower. This leads to some differences in sound propagation, as indicated by the reduced number of cells affected by underwater noise for vessel $D$ compared to vessel $C$, along with the lower noise levels received in those cells. For instance, the cell containing vessel $C$ stores 19.3~dB, whereas the cell containing vessel $D$ records only 11~dB, highlighting a difference of approximately 8~dB. 
These comparisons highlight how a fishing vessel generates more substantial underwater noise than a boat merely sailing, even when the latter has a higher speed, as well as how higher vessel speed contributes to greater underwater sound propagation. 

Finally, we would like to point out that these are only a few examples of the analyses that can be performed using the spatio-temporal dataset of semantic trajectories. For instance, we can focus on vessels equipped with specific fishing gear (i.e., LOTB, SOTB, RAP, and PTM) and determine their impact on the underwater noise level. This fine-grained analysis could help to reveal different pollution degrees of fisheries that, in turn, could constitute a basis to implement specific management actions for these activities. Moreover, we can vary our analysis according to different periods and consider only certain sea areas. For instance, one could focus on protected areas, like the Pomo Pit or the Sole Sanctuary.

\subsection{On model validation}
\label{sec: validation}
Our model has been calibrated by using the real measurements provided by the project SOUNDSCAPE~\cite{SoundscapeProj}, which carried out a continuous acoustic monitoring in the Northern Adriatic Sea between March 2020 and June 2021 at nine stations. As already mentioned, we considered the dataset containing the 60 seconds SPL in the frequencies 63~Hz, 125~Hz, 400~Hz and 4000~Hz. 

Therefore a natural idea for validating our model might consist in comparing the produced results  with the real measurements from the SOUNDSCAPE hydrophones. 
We start by remarking that a direct quantitative comparison is not possible since our model is aimed to estimate the underwater noise generated only by fishing vessels, while the hydrophones recorded the noise generated by all shipping
vessels, including also touristic and commercial vessels. 

\begin{figure}[ht]
  \centering
\includegraphics[width=0.76\linewidth]{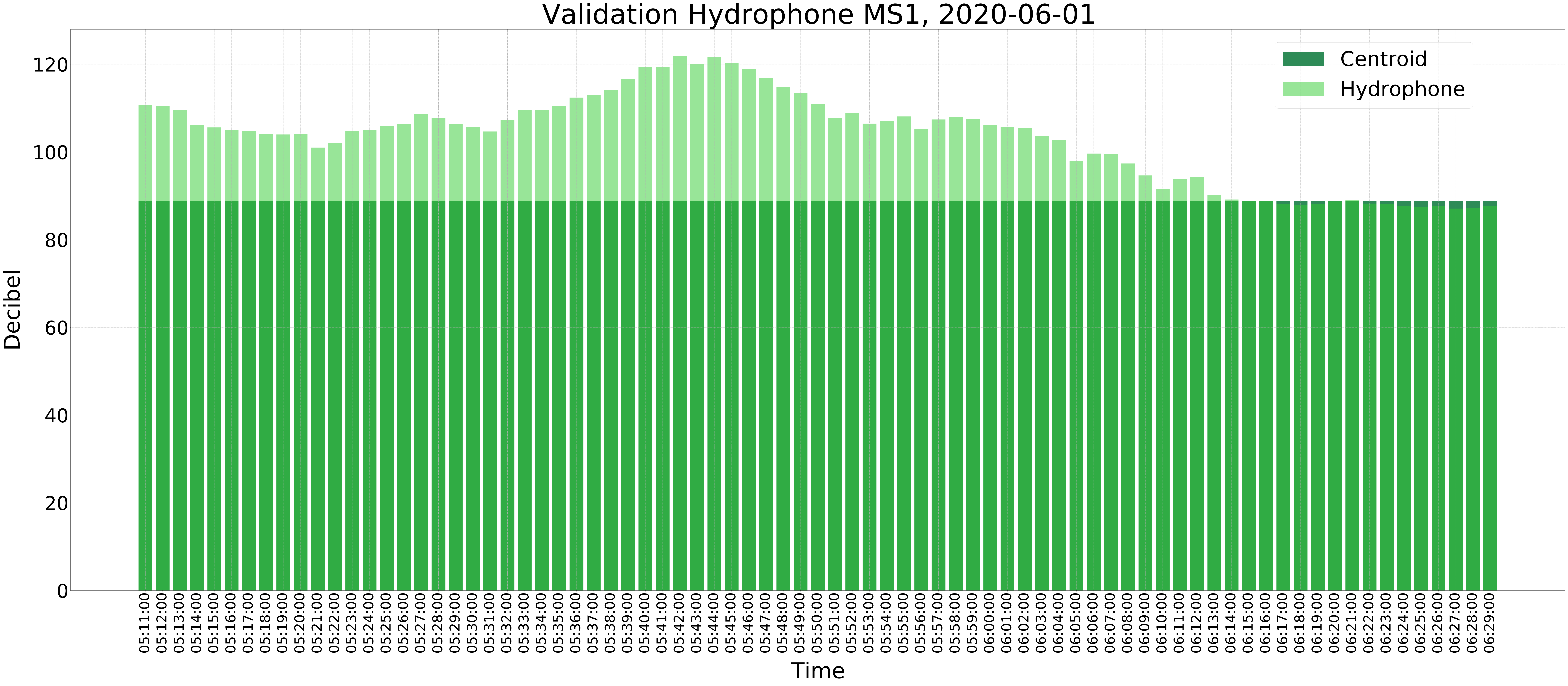}
\caption{Underwater noise measurements at 400~Hz from 5:11~am to 6:29~am on June 1, 2020. For each minute, the underwater noise recorded by hydrophone MS1 (in light green) is represented, as well as the noise recorded by the centroid of the cell in which the hydrophone is located calculated using our model (in dark green).} 
  \label{fig:validationNoBoat}
\end{figure}

Still, some interesting observations can be made, suggesting the adequacy of our model.  In general, given that our model restricts to fishing vessels, we can  only expect that we provide an underestimation of the underwater noise. 
An example of proper underestimation is shown in Figure~\ref{fig:validationNoBoat}. In light green, the recorded values for hydrophone MS1 clearly detects the passage of a vessel between 5:11~am and 6:12~am, after which the measurements return to the ambient noise level (from 6:13~am to 6:29~am). Instead, our model (dark green) remains constantly on the ambient noise level. In fact, no fishing vessel passes in the cell where the hydrophone is located during that time period. 

\begin{figure*}[ht]
\centering
\begin{subfigure}{0.49\textwidth}
    \includegraphics[width=\linewidth]{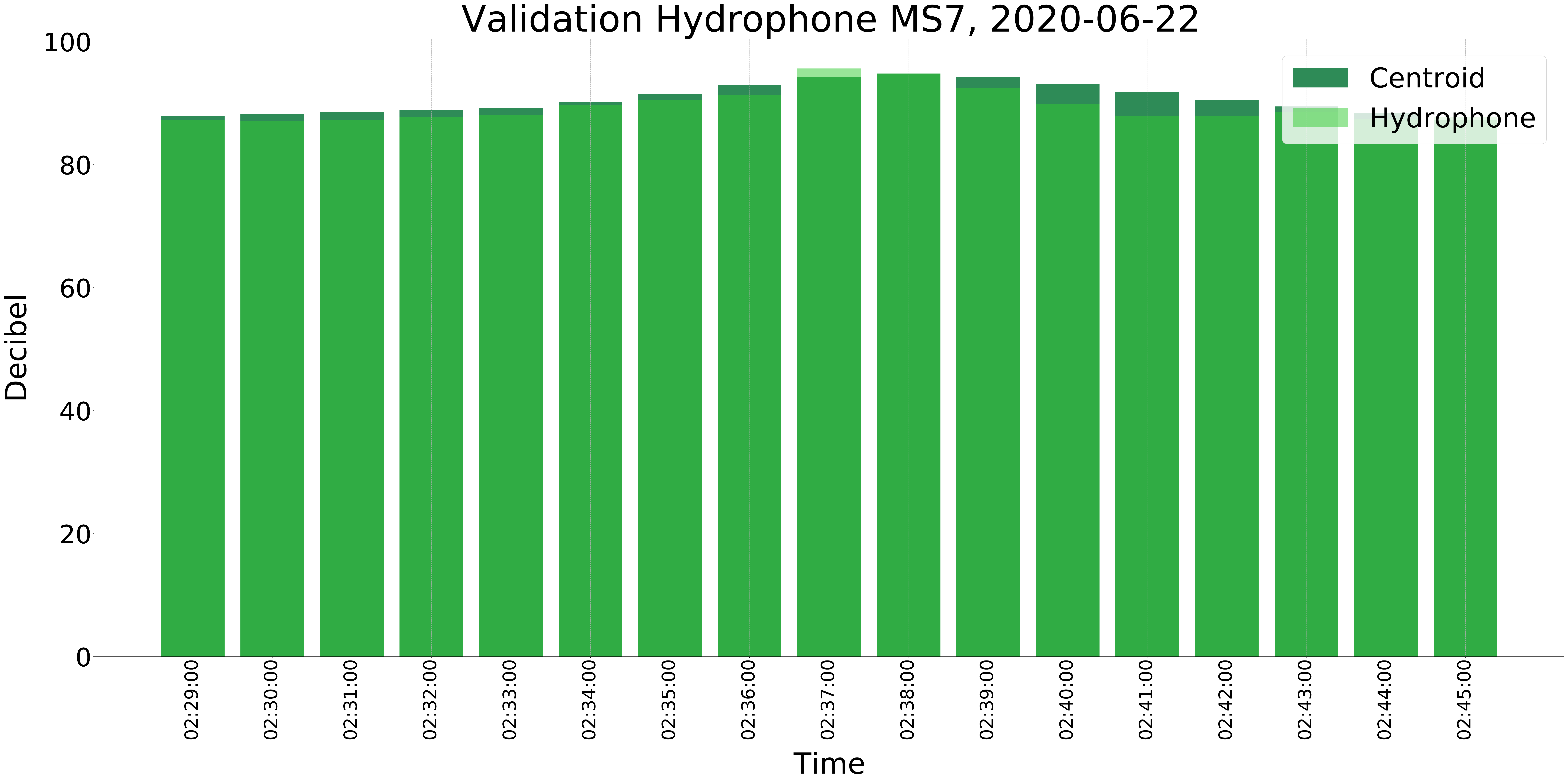}
    \caption{June 22, 2020, from 02:29~am to 02:45~am. Hydrophone MS7 at 125~Hz.} 
    \label{fig:validation63hz}
\end{subfigure}
\hspace{1mm}
\begin{subfigure}{0.49\textwidth}
    \includegraphics[width=\linewidth]{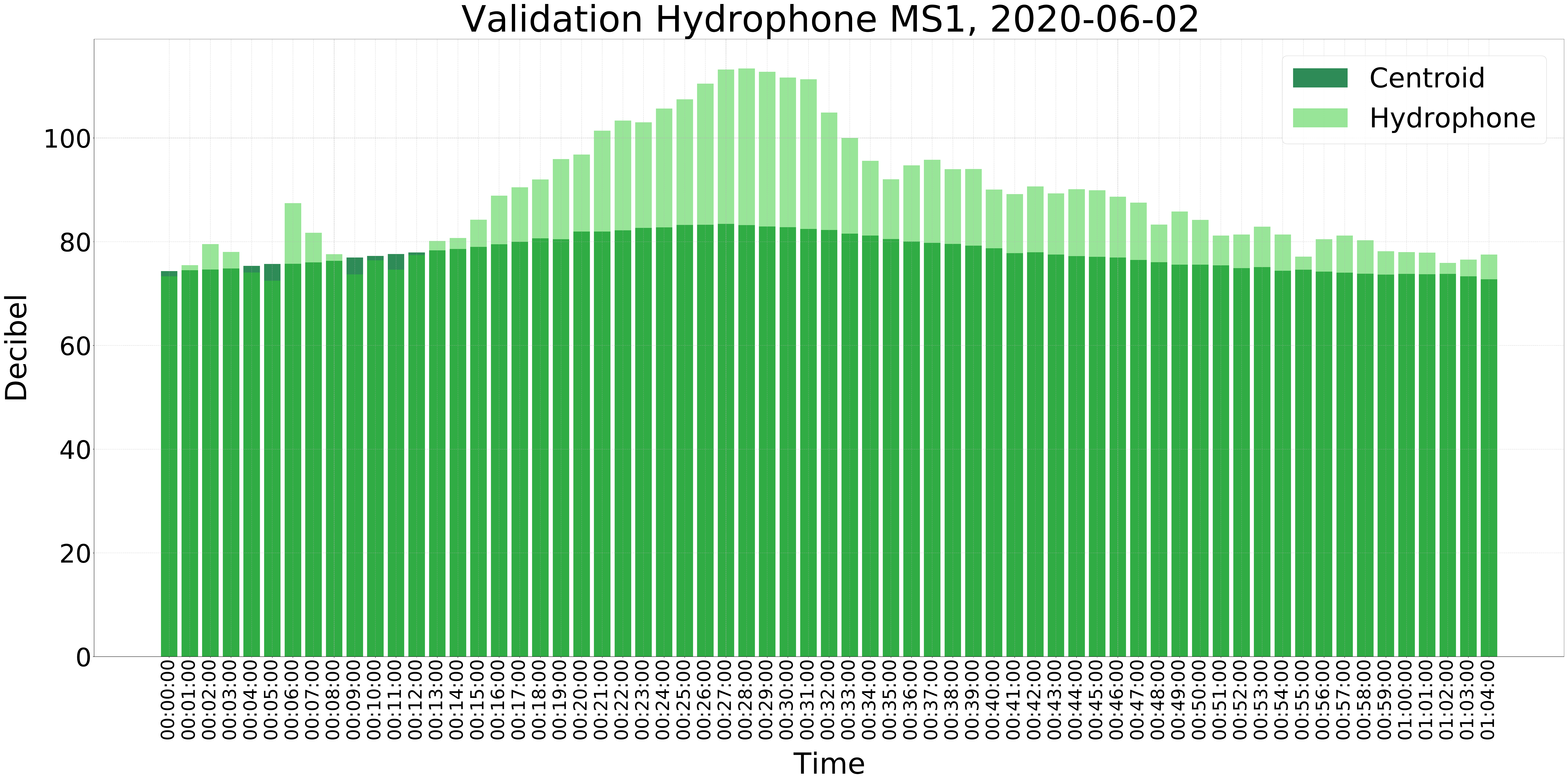}
    \caption{June 2, 2020, from 00:00~am to 1:04~am. Hydrophone MS1 at 125~Hz.} 
    \label{fig:validation63hz_otherVessel}
\end{subfigure}
\caption{Underwater noise measurements at different frequencies taken for each minute that a fishing vessel passes through the cell where a hydrophone is located. For each minute, the underwater noise detected by the hydrophone is represented (in light green), along with that detected by the cell centroid calculated using our model (in dark green).}  
\label{fig:validation}
\end{figure*}

A closer correspondence can be detected when a  fishing vessel is crossing a cell in which a hydrophone is located. In this case, one can observe a similar trend in the underwater noise level
received by the hydrophone compared with the one generated by our model. 
An example is in Figure~\ref{fig:validation63hz} which illustrates minute-by-minute data received by hydrophone MS7 at 125~Hz (in light green) taken on June 22, 2020, from 02:29~am to 02:45~am, compared with the noise levels computed using our model for the virtual listening point located at the centroid of the hydrophone's cell (in dark green). Observe that from 02:29~am the noise level received both at the hydrophone and at the virtual listening point rises, reaching a peak at 02:38~am, before gradually decreasing again until the end of the observations. 
Some slight differences can be possibly explained by recalling that the centroid and the hydrophone do not share identical geographic coordinates (hydrophone MS7 is situated approximately 274.59 meters away from the cell centroid). 
Also, variations in environmental conditions and natural sound sources are always present.
Figure~\ref{fig:validation63hz_otherVessel} instead reveals a difference between our noise estimation and the noise measured by the corresponding hydrophone which extends through almost the entire time period of analysis (around one hour).  This can be interpreted as a situation in which the hydrophone records the noise produced by a fishing vessel passing through the cell (whose trend is replicated in our model) superposed with additional noise produced by non-fishing vessels that are invisible in our model. 

\section{Concluding remarks}
\label{sec: conclusion}
Monitoring the underwater noise pollution due to human activities is an important task for maintaining a healthy marine ecosystem. In this paper, we proposed a framework for the characterisation of underwater noise based on semantic trajectories. 
Starting from the AIS data transmitted by vessels, we reconstructed the  vessels' trajectories and deployed them in a spatio-temporal database. The trajectories were enriched with semantic annotations useful to infer how the noise spreads in the area of interest. 
To estimate the noise generated by vessels, we defined a model for underwater sound propagation based on a combination of   spherical and mode stripping propagation. The model takes into account the bathymetry of the area and the relevant environmental variables, such as sea temperature, pressure  and sea salinity. 

As a case study, we examined the fishing activities in the Northern Adriatic Sea during the year 2020. We implemented the spatio-temporal database of the fishing vessels trajectories using MobilityDB and used the platform to perform some analyses, aiming to showcase the flexibility and expressiveness of our approach. 
We presented also a qualitative validation of our model w.r.t. the SOUNDSCAPE measurements. 
A quantitative validation was not possible because we only consider the AIS data of fishing vessels, so the other marine traffic is invisible to our model. However, for the transits of fishing vessels only, the noise perceived by our virtual listening points seems to be aligned with the one perceived by the real hydrophones.

The proposed approach has many advantages. First, having a spatio-temporal database that stores the noise levels produced by vessels allows to readily obtain noise maps at different time and space granularities, e.g., on a daily basis, or monthly, or seasonally. 
Second, the framework can be used in absence of hydrophones, which can be expensive to install and maintain and cover a limited area. 
In our case study we do calibrate the ambient noise of the model with respect to real measurements from hydrophones, but it is possible to use the model by simply assuming a reasonable value for the ambient noise.
Third, it allows for distinguishing contributions to the underwater noise from different ship types, not only fishing vessels but also commercial or cruise ships, provided that the AIS data are available.

A drawback of our approach is that the boats without an AIS transceiver cannot be modelled. 
It is therefore not possible to estimate their contribution to the total noise, which, as a consequence, could be underestimated. 
This means that areas of the sea showing high values of underwater noise are surely risky for the underwater world, whereas areas that result to be quiet could hide some untracked noise. 

\backmatter




\bmhead{Acknowledgements}
We thank Fabio Pranovi for providing us the AIS data and his valuable knowledge as domain expert. We thank Elisabetta Russo for useful discussions.

This study was funded by the European Union - NextGenerationEU, in the framework of the iNEST - \emph{Interconnected Northeast Innovation Ecosystem} (iNEST ECS\_00000043 – CUP H43C22000540006). 
The second author was supported by the project “Multiscale Analysis of Human and Artificial Trajectories: Models and Applications” funded by the MUR PRIN - grant 2022RB939W.
%
The views and opinions expressed are solely those of the authors and do not necessarily reflect those of the European Union, nor can the European Union be held responsible for them.

\bibliography{sn-bibliography-LONG}
\end{document}